\documentclass[]{gamdie}
\usepackage{graphicx}
\usepackage{amsmath,amssymb,amsfonts}
\usepackage{gensymb}
\usepackage{bm}
\usepackage{upgreek}
\usepackage{wasysym}
\usepackage{bibentry}
\usepackage[T1]{fontenc}
\usepackage{textcomp}
\usepackage{mathptmx}
\usepackage[scaled=0.92]{helvet}
\usepackage{courier}
\newcommand{\eref}[1]{Eq.~\eqref{#1}}
\newcommand{\fref}[1]{Fig.~\ref{#1}}
\newcommand{\upd}{{\ensuremath{\textrm{d}}}}

\DeclareMathOperator\erf{erf}

\begin{document}
\doi{}
\issn{}
\issnp{}
\jvol{}
\jnum{} \jyear{2010} 


\title{Trapping colloids near chemical stripes via critical Casimir forces}

\author{%
Matthias Tr{\"o}ndle$^{a,b}$$^{\ast}$\thanks{$^\ast$Corresponding author.
Email: troendle@mf.mpg.de \vspace{6pt}},
Olga Zvyagolskaya$^{c}$,
Andrea Gambassi$^{d,e}$,
Dominik Vogt$^{c}$,
Ludger Harnau$^{a,b}$,
Clemens Bechinger$^{a,c}$ and
Siegfried Dietrich$^{a,b}$%
\\[6pt]
$^{a}${\em{Max-Planck-Institut f\"{u}r Metallforschung, Heisenbergstr.3, 70569 Stuttgart, Germany}};\\
$^{b}${\em{Institut f\"ur Theoretische und Angewandte Physik, Universit\"at Stuttgart, Pfaffenwaldring 57,     70569 Stuttgart, Germany}};\\
$^{c}${\em{2. Physikalisches Institut, Universit\"at Stuttgart,Pfaffenwaldring 57, 70569 Stuttgart, Germany}};\\
$^{d}${\em{SISSA --- International School for Advanced Studies and INFN, via Bonomea 265, 34136 Trieste, Italy}};\\
$^{e}${\em Laboratoire de Physique Th\'eorique et Hautes Energies, UMR 7589, Universit\'e Pierre et Marie Curie - Paris VI,
4 Place Jussieu, 75252 Paris Cedex 05, France
}
\\[6pt]
  {\centerline{(December 1st, 2010)}}
}

\maketitle
\thispagestyle{empty}

\begin{abstract}
  We  study theoretically and experimentally the solvent-mediated critical Casimir force
  acting on colloidal particles immersed in a binary liquid mixture of water and 2,6-lutidine 
  and close to substrates which are chemically patterned
  with periodically alternating stripes of antagonistic adsorption preferences. 
  These patterns are experimentally realized via microcontact printing.
  Upon approaching the critical demixing point of the solvent, normal and lateral critical Casimir forces
  generate laterally confining effective potentials for the colloids.
  We analyze in detail the rich behavior of the spherical colloids close to such substrates.
  For all patterned substrates we investigated, our measurements of these effective potentials 
  agree with the  corresponding theoretical predictions.
  Since both the directions and the strengths of the critical Casimir forces can be tuned by minute
  temperature changes, this provides a new mechanism for controlling colloids as model systems, opening 
  encouraging perspectives for applications.
  \bigskip
  \begin{keywords}
  Solvation forces, critical Casimir effect, colloidal suspension, chemically patterned substrates
  \end{keywords}
\end{abstract}
\section{Introduction}
%
The structures and the dynamics of colloidal suspensions are to a large extent determined by the effective
interactions among the mesoscopic colloids which are mediated by the solvent  surrounding them.
A rather intriguing example of such solvation forces is the critical Casimir effect \cite{fisher:1978}
which occurs close to a critical point of the solvent.
As a soft-matter analogue of the quantum-electrodynamic Casimir effect \cite{casimir:1948,kardar:1999},
the confinement of the concentration fluctuations within a near-critical binary liquid mixture by the 
surfaces of the colloidal particles results in an attractive or repulsive effective interaction \cite{krech:book,brankov:book,
gambassi:2009conf}.
Upon approaching the bulk critical demixing point of  the mixture at temperature $T=T_c$
the critical fluctuations become long-ranged so that the resulting effective interaction 
among colloids can extend well beyond the molecular scale of the solvent; it may  
act even at distances comparable to the size of the colloids.
The strength and the spatial range of the critical Casimir force depend sensitively on temperature 
because the correlation length $\xi$, which characterizes the extension of the spatial correlations of the critical 
fluctuations and therefore sets the interaction range, diverges as $\xi\simeq \xi_0|(T-T_c)/T_c|^{-\nu}$ upon 
approaching the critical point, $T\rightarrow T_c$, where $\nu\simeq0.63$ is a standard bulk critical 
exponent\footnote{%
Here we quote the value of $\nu$ within the Ising universality class, which is relevant for the 
interpretation of experimental data concerning classical binary 
liquid mixtures.
} 
and $\xi_0$ is a solvent-specific microscopic length.
The critical Casimir effect is \emph{universal} in the sense that all quantities related to it, such 
as the various components of the force and the associated potentials, can be suitably expressed in terms of 
scaling functions which are identical for all those fluids undergoing the same kind of transition, 
irrespective of their molecular details and material properties, which actually enter only
via two independent non-universal amplitudes, such as $\xi_0$, and the value of $T_c$.
\par
Besides some indirect quantitative experimental determinations of the critical Casimir force acting within wetting films
\cite{garcia:9902all,fukuto:2005,ganshin:2006}, its \emph{direct} effect  on a spherical colloid close 
to a substrate and immersed in a critical water-lutidine mixture was measured only recently \cite{hertlein:2008}.
These experiments clearly demonstrated that, upon approaching the critical point,  the critical Casimir force can modify 
the effective interaction between the colloid and the substrate by several multiples of $k_BT$.
Moreover, depending on the preference of the confining surfaces for lutidine or water --- denoted as $(+)$ and $(-)$ effective boundary conditions (BC) for the order parameter, respectively --- the critical 
Casimir force between the colloid and 
the substrate is attractive [for $(\pm,\pm)$ boundary conditions] or repulsive [for $(+,-)$ boundary conditions] \cite{hertlein:2008}.
In addition, these critical Casimir forces depend strongly on the geometrical shape of the confining surfaces
so that topographically structured surfaces experience also a component of the force which is parallel to the walls 
\cite{troendle:2008}, whereas nonspherical particles are subject to a torque \cite{kondrat:2009}.
\par
One can combine these features of the critical Casimir effect by using surfaces which are \emph{chemically patterned}
on the micron-scale in order to obtain also \emph{lateral} critical Casimir forces \cite{sprenger:2006}.
The critical Casimir interaction between a colloid and a substrate which is chemically patterned with stripes of
alternating adsorption preferences results in a laterally confining potential for the colloid, which can be controlled and
reversibly switched on and off by varying the temperature of the near-critical solvent
\cite{soyka:2008,troendle:2009,troendle:2010}.
Recently, such a substrate has been realized by removing via a focused ion beam a regular pattern of equally spaced stripes 
from the hydrophobic HMDS-coating of a glass surface \cite{soyka:2008}. 
The effective interaction between this chemically patterned substrate and a colloidal particle has then been measured by 
monitoring via digital video microscopy the behaviour of a dilute solution of colloidal particles as the demixing
critical point of the binary mixture of water and lutidine is approached \cite{soyka:2008}. 
In comparing these data with the corresponding theoretical predictions \cite{troendle:2009}
it turned out that the critical Casimir potential resulting from a chemical pattern depends rather strongly on its geometrical 
details and it actually provides a sensitive tool to probe these microscopic features, 
which might not be easily accessible otherwise \cite{troendle:2009}.
Indeed agreement between the theoretical predictions of Ref.~\cite{troendle:2009} and the experimental 
data of Ref.~\cite{soyka:2008} for the spatially varying effective potential was found only by assuming 
that the chemical steps between two subsequent stripes within the pattern were not microscopically sharp due 
to fluctuations inherent to the preparation process of the structures \cite{troendle:2009}. 
However, in spite of this improvement in the comparison between the \emph{shapes} of the theoretical and 
experimental potentials, the analysis of the temperature dependence of the corresponding potential \emph{depths}  
renders an amplitude  $\xi_0$ of the correlation length $\xi$  (treated as a fitting parameter) which is about 
twice as large \cite{troendle:2009} as the one known from independent previous measurements. 
This persisting discrepancy, together with the necessity to test independently the aforementioned assumption used in 
Ref.~\cite{troendle:2009}, calls for a more detailed experimental study of the critical Casimir force acting 
in the presence of patterned substrates.
\par
Colloidal suspensions are not only relevant as model systems in soft-matter physics but they may find also 
applications in integrated micro- and nano-devices.
Since critical Casimir forces provide a tool to exert active control over the strength and the direction of 
interactions between colloids in such devices, a thorough quantitative understanding of the underlying physics is essential.
Recently, it was theoretically predicted that for suitable geometrical parameters colloids can levitate in a stable position
above chemically patterned substrates \cite{troendle:2010}.
Such a ``critical Casimir levitation'' may even help in overcoming the problem of 
stiction which hampers the functioning of currently available micro- and nano-electromechanical devices.
Therefore, in order to benefit from this wide range of possibilities, a theoretical understanding of the 
phenomenon and the experimental demonstration of its applicability are essential.
\par
Here, we report on experimental studies of the critical Casimir interaction between a colloid and substrates which are chemically
structured by microcontact printing of alkane\-thiols. 
We shall see below that, in the present context  the substrates prepared by this technique turn out to exhibit higher 
chemical and geometrical resolution than those prepared via focused ion beam, which was used in previous experiments \cite{soyka:2008}.
A dilute suspension of hydrophilic spherical colloids of radius $R$ which impose $(-)$ boundary conditions
to a near-critical water-lutidine mixture is exposed to patterned substrates
consisting of alternating and periodically repeating stripes of width $L_\pm$ with boundary conditions $(\pm)$ so that the
periodicity is $P=L_++L_-$ along the $x$-direction, whereas the pattern is translationally invariant along the 
orthogonal $y$-direction.
In order to extend previous investigations~\cite{soyka:2008,troendle:2009} to a wider range of chemical patterns,  
we consider here sequences of stripes with various periodicities $P$ and stripe widths $L_-$.
In particular this allows us to investigate the case in which the effects of neighboring
chemical steps interfere, which strongly affects the resulting forces acting on the colloid.
As in Ref.~\cite{soyka:2008} we measure the laterally varying equilibrium spatial distribution of 
the colloids by digital video microscopy and define an effective potential on the basis of the equilibrium number 
density of the colloids projected along the vertical $z$-direction (orthogonal to $x$ and $y$).
We analyze theoretically this potential and the ensuing surface-to-surface distance $z$ of the colloid from the 
substrate and compare our findings with the experimental data. 
Our theoretical analysis provides accurate information also about the vertical probability distribution of 
the positions of the colloidal particles  which, however, cannot be resolved by the kind of video microscopy used  
in the current experiment.
For all stripe widths and temperatures studied we find quantitative agreement between the measurements and 
the theoretical predictions.
Both the shapes and the depths of the measured potentials agree very well with the theoretical analysis.
Moreover, the uncertainty of the actual position of the subsequent chemical steps is 
significantly reduced by using the present microcontact printing in order to structure the substrate.
The amplitude $\xi_0$ of the bulk correlation length, which we estimate by comparing  the depth of the measured
potentials with the theoretical predictions, turns out to be in quantitative agreement with previous, independent estimates.
Thus, our detailed comparison of experimental data and theory demonstrates that indeed the full benefits
of the critical Casimir force  can be reaped reliably  and  may be utilized 
in potential applications, opening a new route for using colloids as model systems or in micro- and nano-devices.
\section{Theory}
%
\subsection{Critical Casimir potential\label{sec:casimir}}
%
According to renormalization group theory, in the vicinity of the critical point at $T=T_c$ 
the normal and lateral critical Casimir forces as well as the corresponding potential can
be described by \emph{universal scaling functions} (see, e.g.,
Refs.~\cite{krech:book,brankov:book,krech:9192all}).
These scaling functions are universal in the sense that they do not depend on 
molecular details but only on the gross features of the system and of the confining surfaces.
These gross features, among which there are the spatial dimension $d$, the symmetries of the system, and the geometry 
of the boundaries, determine the so-called bulk \cite{krech:book,brankov:book} and surface  
\cite{binder:1983,diehl:1986,diehl:1997,krech:book} universality classes of the associated critical point.
The relevant thermodynamic properties which emerge upon approaching the point of the continuous phase transition can 
be understood and analyzed in terms of the fluctuations of the so-called order parameter $\phi$
of the phase transition.
For the consolute point of phase segregation of a binary liquid mixture $\phi$ is given by the
difference between the local and the mean concentration of one of the two components of the mixture;
thus, it is a scalar quantity and the bulk universality class is of the so-called Ising type (see, e.g.,
Refs.~\cite{gambassi:2009,hertlein:2008}).
For classical binary liquid mixtures, the surfaces generically exhibit preferential adsorption for one of the
two species of the mixture, which results in a local enhancement of the order parameter $\phi$ close to the
surface.
This enhancement is effectively described by symmetry-breaking surface fields 
\cite{binder:1983,diehl:1986,diehl:1997,krech:book} and it is usually denoted by $(+)$ and $(-)$ boundary 
conditions (BC) corresponding to having a preference for $\phi>0$ and $\phi<0$, respectively, at the surface. 
For the water-lutidine mixture we are interested in, one conventionally indicates the preferential adsorption for 
lutidine and water as $(+)$ and $(-)$ BC,  respectively (see also
Refs.~\cite{hertlein:2008,gambassi:2009}).
In the experimental setup described below, the lower critical demixing point of the water-lutidine mixture
is always approached from the homogeneous (mixed) phase at temperatures $T<T_c$ upon increasing $T$ at 
\emph{fixed critical composition} of the mixture.
From the experimental point of view it is rather difficult to quantify the strength of the adsorption preference 
exhibited by the different portions of the surfaces.
Therefore the comparison with theoretical predictions requires  assumptions, which can be verified a posteriori. 
In the present case we have qualitative experimental evidence that the chemical treatment of the surfaces results 
in rather pronounced adsorption preferences and therefore we shall assume that  the surfaces are characterized by the 
so-called \emph{strong critical adsorption} fixed point \cite{binder:1983,diehl:1986}.
We consider here neither the opposite case of weak adsorption at the surfaces (see, e.g., Refs.~\cite{mohry:2010,nellen:2009}) 
nor  the effects due to off-critical compositions of the mixture, which might lead to a bridging transition 
analogous to capillary condensation (see, e.g., Refs.~\cite{gambassi:2009,schlesener:2003,drzewinski:2000,
maciolek:2001,evans:1990,bieker:1998,bauer:2000}).
\par
For the film geometry --- in which the binary liquid mixture is confined between two parallel, planar, and macroscopically
extended walls at a distance $L$ --- the normal critical Casimir force per unit area is given by \cite{krech:9192all}
\begin{equation}
  \label{eq:planar-force}
  f_{(+,\pm)}(L,T)=k_BT \frac{1}{L^d}k_{(+,\pm)}(L/\xi),
\end{equation}
where the index $(+,\pm)$ denotes the combination of BC $(+)$ and $(-)$ at the two walls.
(In the absence of a symmetry-breaking bulk field within the 
film, which would correspond to an off-critical concentration of the mixture, one has $f_{(-,-)} = f_{(+,+)}$.)
The universal scaling functions $k_{(+,\pm)}$ depend only on the scaled variable $L/\xi$, i.e., on the
film thickness $L$ in units of the bulk correlation length $\xi$.
The functions $k_{(+,\pm)}$ have been calculated exactly in $d=2$ \cite{evans:1994}, for $d<4$
using perturbative field-theoretical methods \cite{krech:1997} or  effective theories \cite{borjan:2008},  and 
in $d=3$ numerically via Monte Carlo simulations \cite{vasilyev:2007,vasilyev:2009,hasenbusch:2010}. 
At present, a quantitatively reliable theoretical determination of both $k_{(+,+)}$ and $k_{(+,-)}$ in 
$d=3$ is provided only by Monte Carlo simulations.
For a chemically structured wall opposing another planar wall forming a film geometry, the critical Casimir force
has been obtained within mean-field theory (MFT) \cite{sprenger:2006} and from Monte Carlo 
simulations \cite{parisen:2010,parisen:2010a}.
\par
For a spherical colloid with $(-)$ BC opposite to a chemically patterned substrate with alternating $(+)$ and $(-)$ 
BC the critical Casimir potential has been calculated both numerically within full mean-field theory (corresponding
to $d=4$) and by resorting to the so-called Derjaguin approximation exploiting the full knowledge of 
$k_{(+,\pm)}$ in the film geometry for $d=3$ \cite{troendle:2010}.
Accordingly, the critical Casimir force and the critical Casimir potential for a colloid close to a chemically
patterned wall can be expressed in terms of universal scaling functions which take the geometry into account and
which depend on the following scaled quantities:
\begin{itemize}
\setlength{\itemindent}{0cm}
  \item {$\Theta=z/\xi$, where $z$ is the surface-to-surface distance between the colloid and the substrate,}
  \item {$\Delta=z/R$, where $R$ is the radius of the spherical colloid,}
  \item {$\Xi=x/\sqrt{Rz}$, where $x$ is the lateral coordinate of the center of the colloid
        such that $x=0$ corresponds to the colloid being located opposite to the center of a stripe with $(-)$ BC,}
  \item {$\Pi=P/\sqrt{Rz}$, where $P$ is the periodicity of the stripe pattern along the $x$-direction,}
  \item {$\lambda=L_-/P$, where $L_-$ is the width of the stripes with $(-)$ BC.}
\end{itemize}
In particular in $d=3$ the critical Casimir potential $\Phi_{\rm C}$ can be written as \cite{troendle:2010}
\begin{equation}
  \label{eq:phi}
  \Phi_{\rm C}(L_-,P,x,z,R,T)=k_BT\frac{R}{z} \vartheta(\lambda,\Pi,\Xi,\Theta,\Delta),
\end{equation}
where $\vartheta$ is a universal scaling function.
The realization of the present geometrical setup of a sphere facing a plane still represents a challenge for 
lattice based Monte Carlo simulations. 
Accordingly, up to now it is not possible to obtain accurate numerical data for this three-dimensional geometry and 
therefore one has to rely on the Derjaguin approximation (DA), based on the assumption 
of additivity, in order to calculate approximately the critical Casimir potential (see Ref.~\cite{troendle:2010}).
Within the Derjaguin approximation, the scaling function $\vartheta$ [\eref{eq:phi}] can be expressed in terms of
the known scaling functions $k_{(\pm,-)}$ for the film geometry \cite{troendle:2010}:  
\begin{equation}
  \label{eq:-omega}
  \vartheta(\lambda,\Pi,\Xi,\Theta,\Delta\to0)=
  \frac{\vartheta_{(+,-)}(\Theta)+\vartheta_{(-,-)}(\Theta)}{2}+
  \frac{\vartheta_{(+,-)}(\Theta)-\vartheta_{(-,-)}(\Theta)}{2}
  \omega(\lambda,\Pi,\Xi,\Theta),
\end{equation}
where
\begin{equation}
  \label{eq:derjaguinpot}
  \vartheta_{(\pm,-)}(\Theta)=
  2\pi\int_{1}^{\infty}\upd \beta\left( \beta-1 \right)\beta^{-3}k_{(\pm,-)}(\beta\Theta)
\end{equation}
are the scaling functions of the critical Casimir potential of a colloid in front of
a homogeneous wall as calculated within the Derjaguin approximation \cite{hertlein:2008},
and
\begin{equation}
  \label{eq:omega}
  \omega(\lambda,\Pi,\Xi,\Theta)=
    1+\sum_{n=-\infty}^{\infty} \left\{\omega_{\rm s}(\Xi+\Pi(n+\tfrac{\lambda}{2}),\Theta)
    -\omega_{\rm s}(\Xi+\Pi(n-\tfrac{\lambda}{2}),\Theta)\right\}
\end{equation}
with $\omega_{\rm s}$ given by
\begin{equation}
  \label{eq:step-omega-da}
  \omega_{\rm s}(\Xi\gtrless0,\Theta)=
  \mp1\pm
  \frac{\Xi^4
  \int_{1}^{\infty}\upd s
  \frac{
  s\arccos\left( s^{-1/2}\right)-\sqrt{s-1}}
  {(1+\Xi^2s/2)^{d}}
  \Delta k
  \left( \Theta[1+\Xi^2 s/2] \right)}
  {
  \vartheta_{(+,-)}(\Theta)-\vartheta_{(-,-)}(\Theta)
},
\end{equation}
where $\Delta k(\Theta) = k_{(+,-)}(\Theta) - k_{(-,-)}(\Theta)$.
For $\Theta=0$ and $\Theta\gg1$ there are analytic expressions for $\omega_{\rm s}$ 
(see Eqs.~(6) and (7) in Ref.~\cite{troendle:2009}, respectively).
The accuracy of the Derjaguin approximation has been checked numerically within mean-field 
theory and it has turned out that this approximation describes quantitatively the actual behavior 
of those numerical data which correspond to $\Theta\gtrsim4$ for $\Delta\lesssim1$ and $\Pi\gtrsim0.5$,
as well as to $0\le\Theta\lesssim4$ with $\Delta\lesssim0.3$ and $\Pi\gtrsim2$ \cite{troendle:2010}.
In the experiments discussed further below the corresponding values of $\Theta$, $\Delta$, and $\Pi$ 
vary within these ranges so that, assuming that the previous quantitative conclusions  
extend to $d=3$, we expect the Derjaguin approximation to be quantitatively reliable.
\subsection{Background forces\label{sec:allforces}}
%
In addition to the critical Casimir force due to the critical fluctuations of the solvent, 
the colloidal particles of the suspension are subjected to additional effective forces which are characterized 
by a smooth and rather mild dependence on temperature. 
Typical background forces acting within the colloidal suspensions of present interest are due to  
(screened) electrostatic and dispersion interactions and to the gravitational field. 
In a first approximation, which neglects possible mutual influences of these forces,  their total potential 
is given by the sum of the corresponding contributions: (i)  the electrostatic potential $\Phi_{\rm el}$,
(ii) the gravitational potential $\Phi_{\rm g}$, 
(iii) the van der Waals interaction $\Phi_{\rm vdW}$, and (iv) the effective critical Casimir potential  $\Phi_{\rm C}$.
\par
\textit{Electrostatics} --- This interaction originates from the fact that, due to the formation of charge 
double-layers, the surface of the colloids and of the substrate acquire a surface charge once immersed in the 
liquid solvent. 
As a result,  the polystyrene colloids of radius $R=1.2\mu$m and $(-)$ BC
immersed in water-lutidine mixtures experience an electrostatic repulsion from the substrate.
The screened electrostatic potential of the colloid at a surface-to-surface distance $z$ from
a \emph{homogeneous} substrate with $(\pm)$ BC is well approximated by
\begin{equation}
  \label{eq:elhomog}
  \Phi_{ {\rm el},\pm}(z)/(k_BT)=\exp\{-\kappa(z-z_{0}^{\pm})\},
\end{equation}
where $\kappa^{-1}$ is the screening length and $z_0^\pm$ describes the strength of the electrostatic 
repulsion from the substrate with $(\pm)$ BC. 
Although the values of these parameters are determined by the surface charge of the colloid, the dielectric 
constant of the mixture etc. \cite{israelachvili:book}, here they will be treated as fitting parameters of 
the actual experimental data for $\Phi_{ {\rm el},\pm}$.
In experimental conditions similar to the present ones as far as the mixture and the colloids are concerned, 
one finds $\kappa^{-1}\simeq12$nm \cite{hertlein:2008,nellen:2009,gambassi:2009} 
and $z_0^\pm\simeq0.1-0.2\mu$m \cite{hertlein:2008,nellen:2009,gambassi:2009,troendle:2009} as typical values 
at $T\simeq T_c\simeq 307K$.
In view of a possible difference between $z_0^+$ and $z_0^-$, the resulting electrostatic potential of a colloid close to a
\emph{patterned} substrate such as the one described above depends on the lateral 
position $x$ of the colloid. 
Such a dependence can be accounted for theoretically within the Derjaguin approximation (by repeating the calculations described 
in Appendices A.2, B, and C in Ref.~\cite{troendle:2010} for  $\Phi_{\rm C}$), which is expected to be particularly 
accurate due to the exponential decay of  $\Phi_{ {\rm el},\pm}$ in \eref{eq:elhomog} as a function of the distance $z$ 
from the substrate.
For this electrostatic potential $ \Phi_{\rm el}(x,z)$ one therefore finds (for a colloid facing the center 
of a $(-)$ stripe at $x=0$)
\begin{equation}
  \label{eq:phiel}
  \Phi_{\rm el}(x,z)=\frac{\Phi_{ {\rm el},+}(z)+\Phi_{ {\rm el},-}(z)}{2}
                    +\frac{\Phi_{ {\rm el},+}(z)-\Phi_{ {\rm el},-}(z)}{2}
                    \Omega(x),
\end{equation}
where
\begin{equation}
  \label{eq:elomega}
  \Omega(x)=1+\sum_{n=-\infty}^\infty\left\{\erf([x+nP-L_-/2]/\Lambda)-\erf([x+nP+L_-/2]/\Lambda)\right\},
\end{equation}
with $\Lambda=\sqrt{2 R\kappa^{-1}}\simeq0.17\mu$m.
$\Omega$ and $\Phi_{\rm el}$ depend on the geometric parameters $L_-$ and $P$ describing the pattern.
\par
\textit{Dispersion forces} ---
For the particular choice of materials and conditions used in the present experiments, van der Waals forces
turn out to be negligible compared with the other contributions \cite{hertlein:2008,gambassi:2009,dantchev:2007,troendle:2009}. 
In addition, the temperature dependence of the dielectric permittivity of the water-lutidine mixture is negligible 
(see footnote 4 in Ref.~\cite{troendle:2009}) and therefore no significant changes are expected to occur in $\Phi_{\rm vdW}$ 
and $\Phi_{\rm el}$ within the range of temperatures explored experimentally.
\par
\textit{Gravity} ---
Due to buoyancy the colloid immersed in the solvent above the patterned substrate experiences a gravitational
potential given by
\begin{equation}
  \label{eq:phig}
  \Phi_{\rm g}(z)=(\rho_{\rm PS}-\rho_{\rm WL})g\frac{4\pi}{3}R^3\;z \equiv G\,z,
\end{equation}
where $\rho_{\rm PS}\simeq1.055$g/cm$^3$ and  $\rho_{\rm WL}\simeq0.988$g/cm$^3$ \cite{jayalakshmi:1994}
are the mass densities of the polystyrene colloid and of the water-lutidine mixture at the critical composition and near $T_c$, 
respectively, and
$g\simeq9.81$m/s$^2$ is the gravitational acceleration.
Accordingly, at $T\simeq307$K one has $G\simeq1.12 k_BT / \mu$m and therefore, compared to the other contributions, it
turns out that the gravitational potential depends rather mildly on the distance $z$ because it varies only over a few $k_BT$ 
on the relevant length scale of a few microns.
The expression in \eref{eq:phig} assumes that the colloidal particle of mass density $\rho_{\rm PS}$ is floating in 
a homogeneous medium of mass density $\rho_{\rm WL}$.
However, the laterally varying adsorption preferences of the substrate induce the formation of alternating water-rich 
or lutidine-rich regions close to the surface of the patterned substrate, which laterally alter the resulting mass density 
of the solvent  as a consequence of  water and lutidine having different mass densities. 
This implies that the effective gravitational constant $G$ acquires a dependence on $x$.  
In addition, the preferential adsorption of the colloid, with the ensuing formation of an adsorption profile around it, 
can lead to a modification of the effective density $\rho_{\rm PS}$ of the colloid itself.
However, on the basis of our estimates, we expect all these effects to be negligible for the present experimental 
conditions \cite{troendle:2009}.
\subsection{Total potential}
\begin{figure}
  \hspace*{-20mm}
  \begin{minipage}{170mm}
  \subfigure[\large $\xi=5$nm]{
  \resizebox*{55mm}{!}{\includegraphics[trim= 24pt 12pt 12pt 25pt,clip]{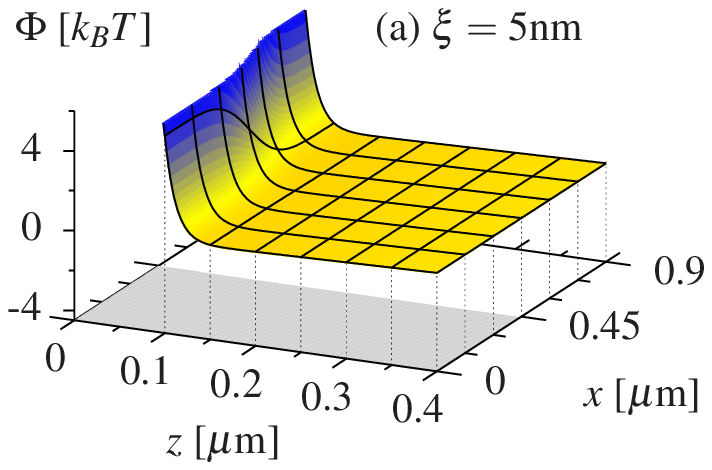}}}
  \subfigure[\large $\xi=22$nm]{
  \resizebox*{55mm}{!}{\includegraphics[trim= 24pt 12pt 12pt 25pt,clip]{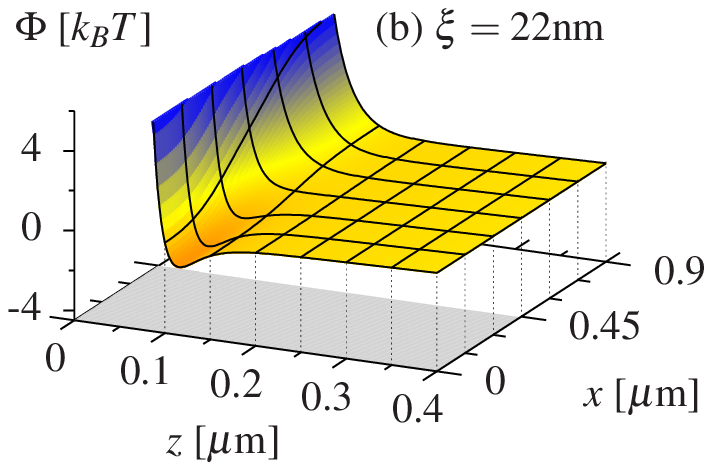}}}
  \subfigure[\large $\xi=26$nm]{
  \resizebox*{60mm}{!}{\includegraphics[trim= 24pt 12pt 0pt 25pt,clip]{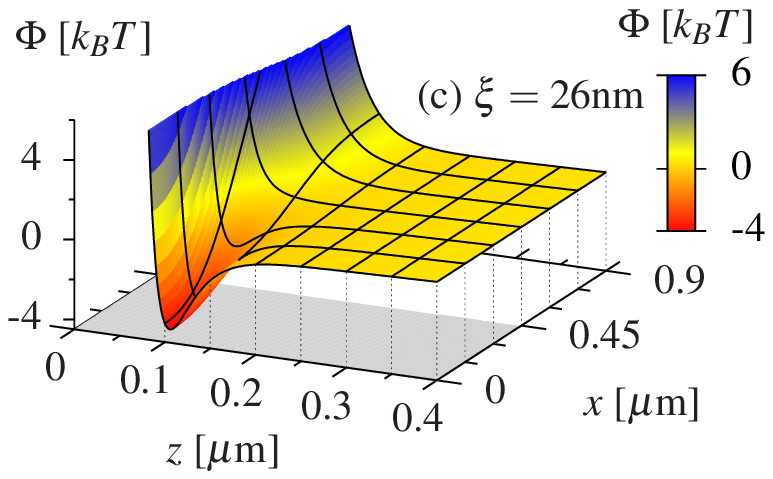}}}
\end{minipage}
  \caption{%
    Total potential $\Phi(x,z,T)$ of a colloid with radius $R=1.2\mu$m opposite to a chemically patterned substrate
    with $L_-=0.9\mu$m and $P=1.8\mu$m for three temperatures corresponding to (a) $\xi=5$nm, (b) $\xi=22$nm, and
    (c) $\xi=26$nm, respectively.
    The electrostatic potential corresponds to $z_0^-=0.12\mu$m, $z_0^+=0.08\mu$m, and $\kappa^{-1}=12$nm 
    (see the main text for details).
    At small separations $z\lesssim z_0^\pm$, the colloid is strongly repelled from the substrate due to electrostatics.
    The effective gravitational potential associated with buoyancy is rather flat, with a spatial slope
    $G\simeq1.12 k_BT / \mu$m. 
    Accordingly, the surface-to-surface particle-substrate distance $z$ exhibits large thermal fluctuations of the 
    order of $k_BT/G \simeq 1\mu$m as long as $\xi$ is small.
    Upon approaching the critical point [from (a) to (c)], $\xi$ increases and a deep, local potential minimum arises rapidly
    as a function of $\xi$ above that part of the substrate (indicated by the shaded area) with the same preferential adsorption 
    as the colloid.
    The colloid is eventually confined in this deep potential well at a distance $z\simeq z_0^-$ with almost no vertical  fluctuations.
    The black lines correspond to cuts through the potential surface at constant values of $z$ and $x$, respectively.
  }
  \label{fig:3d}
\end{figure}

%
The total potential $\Phi$ of the sum of the forces acting on the colloid is given by
\begin{equation}
  \label{eq:phitot}
  \Phi(x,z,T)=\Phi_{\rm C}(L_-,P,x,z,R,T)+\Phi_{\rm el}(x,z)+\Phi_{\rm g}(z),
\end{equation}
where the theoretical expressions for the individual contributions are given by Eqs.~\eqref{eq:phi}, 
\eqref{eq:phiel}, and \eqref{eq:phig}. 
Here and in the following we do not indicate the dependence of $\Phi$ on $L_-$, $P$, and $R$, because the values 
of these parameters are fixed for each individual experiment.
Figure~\ref{fig:3d} shows the total potential $\Phi(x,z,T)$ of a single colloid with $(-)$ BC opposite to
a chemically patterned substrate, as a function of both $x$ and $z$ and for three values of the temperature $T$ 
close to the critical value $T_c$. 
These three values correspond to different correlation lengths $\xi$, as indicated in the figure. 
The gray area in the $x$-$z$ plane indicates the vertical projection of the stripe with $(-)$ BC, the center of which 
corresponds to $x=0$. 
In \fref{fig:3d} the white part of the $x$-$z$ plane corresponds to the projection of the stripe with $(+)$ BC, the center of which is 
located at $x = P/2 = 0.9\mu$m.
The potential of the forces acting on the colloid is translationally invariant along the $y$-direction
which is not shown in \fref{fig:3d}.
$\Phi(x,z,T)$ in \fref{fig:3d} has been calculated by using geometrical and interaction parameters which mimic the actual 
experimental conditions and by using values of the correlation length  $\xi$ which are experimentally available.
As anticipated above, \fref{fig:3d} clearly shows that the gravitational tail of the potential, which characterizes 
$\Phi(x,z,T)$ at large values of $z$,  is indeed rather flat on the scales of $k_BT$ and of hundreds of nm. 
As a consequence of thermal fluctuations --- which cause the colloid to explore a region of space within which 
the total potential $\Phi$ differs from its minimum by a few $k_BT$ ---  the particle is expected to display large 
fluctuations $\Delta z \simeq k_BT/G$ along the vertical direction.
At small particle-substrate distances electrostatic forces are responsible for the strong repulsion of the colloid from
the substrate.
In \fref{fig:3d} we allowed for a lateral inhomogeneity of the electrostatic potential (i.e., $z_0^+\neq z_0^-$), 
which might occur due to different surface charges on the different stripes. 
This is clearly visible in panel (a) of \fref{fig:3d}, which corresponds to a rather small value of the correlation 
length $\xi$ so that, within the range $z\simeq0.1\mu$m \fref{fig:3d}(a) refers to, the contributions of the critical 
Casimir force are negligible. 
In this case, the $x$-dependent electrostatic contribution dominates at small values of $z$, whereas the laterally 
homogeneous gravitational potential dominates at larger distances.
However, upon approaching the critical point [Figs.~\ref{fig:3d}(b) and \ref{fig:3d}(c)], the correlation length $\xi$ 
increases and the critical Casimir force  acting on the colloid builds up;
it is repulsive within the region with $(+)$ BC whereas it is attractive within the region with $(-)$ BC. 
In the latter case the behavior of the colloid is eventually determined by the electrostatic repulsion and the 
attractive critical Casimir force, whereas in the former case it is determined by gravitation and the repulsive Casimir effect.
Above a certain threshold value of $\xi$, which depends on the specific choice of the various geometrical and physical 
parameters, a very deep and steep potential well develops rapidly close to the stripe with $(-)$ BC,
which therefore confines the vertical motion of the colloid at much smaller values of $z$ than before with very limited
thermal fluctuations of the particle-wall distance. 
In contrast, the vertical repulsive critical Casimir force, which the colloid experiences above the stripe with $(+)$ BC, 
pushes it further away from the surface, but the corresponding fluctuations of the vertical position $z$ 
(still limited only by the gravitational tail) are not significantly affected.
Thus, the full  theoretical analysis of the various forces at play reveals a rather interesting energy landscape which is strongly
temperature dependent.
\subsection{Measured potential}
In order to measure experimentally the total potential $\Phi$ of the forces acting on a colloid a very effective 
approach consists in monitoring  the equilibrium Brownian motion of a \emph{single} particle and inferring 
from the sampled probability distribution function $\bar{P}(x,y,z) \propto \exp(-\Phi(x,z,T)/(k_BT))$  the potential as 
$\Phi/(k_BT) = - \ln \bar{P} + \textit{const}$, where $x$ and $y$ are the lateral coordinates of the projection of the colloid 
center onto the substrate surface.
This approach forms the basis of total internal reflection microscopy, which has been 
used in Refs.~\cite{hertlein:2008,nellen:2009,gambassi:2009} to study  critical Casimir forces.
Alternatively, one can study a colloidal \emph{suspension} which is sufficiently dilute so that the inter-particle
interaction is negligible.
In this case, the mean equilibrium number density $\rho(x,y,z)$ of the colloids at position $(x,y,z)$ is proportional 
to the single-colloid probability distribution function $\bar{P}$ and therefore it is again given by 
$\rho(x,y,z)\propto\exp(-\Phi(x,z,T)/k_BT)$.
In the experimental setup described below, the positions of the centers of the colloids are monitored via a digital video 
camera positioned below the substrate. 
Accordingly, the surface-to-surface distances $z$ of the colloids from the substrate are not resolved and the camera records  
only the projected number density $\rho_{\rm P}(x,y) \equiv \int_0^\infty\upd\,z\; \rho(x,y,z)$.
Due to the translational invariance of the chemical pattern along the $y$ direction of length $l\gg P,R$, the density 
$\rho_{\rm P}(x,y)$ can be conveniently projected further onto the $x$-axis, resulting in an effective  
number density $\hat{\rho}(x)=l^{-1}\int_0^l\upd y\; \rho_{\rm P}(x,y) = l^{-1}\int_0^l\upd y\int_0^\infty\upd\,z\; \rho(x,y,z)$,
which depends only on $x$.
This projection increases the statistics and therefore the accuracy with which this projected density can be determined 
experimentally.
Subsequently, an effective potential $\hat{V}(x)$ (up to an irrelevant 
additive constant) can be associated with $\hat{\rho}(x)$ such that $\hat{\rho}(x)\propto\exp(-\hat{V}(x)/(k_BT))$. 
(Note that due to the thermal fluctuations of the colloids along the vertical direction, even if one knows the average distance 
$z_{\rm avg}(x)$ of the colloid from the substrate at a certain lateral position $x$, the effective potential $\hat{V}(x)$ 
is not simply given by $\Phi(x,z_{\rm avg}(x),T) + \textit{const}$, as it was implicitly assumed in Ref.~\cite{soyka:2008}.)
The measured potential
\begin{equation}
    \label{eq:vhat}
    \delta\hat{V}(x)=\hat{V}(x)-\hat{V}(P/2)=-k_BT\ln(\hat{\rho}(x)/\hat{\rho}(P/2))
\end{equation}
is eventually defined such that it vanishes for a colloid opposite to the center of a repulsive $(+)$ stripe at
$x=P/2$.
We emphasize that $\delta\hat{V}(x)$ contains universal ingredients stemming from the scaling function associated with
$\Phi_{\rm C}$ (see \eref{eq:phitot}) as well as nonuniversal contributions due to $\Phi_{\rm el}$ and $\Phi_{\rm g}$.
\subsection{Non-ideal stripe patterns\label{sec:non-ideal}}
Due to the preparation process (see below) the actual position $x=x_s(y)$ of each chemical step separating
two adjacent stripes might vary smoothly along the $y$-axis. 
This variation affects the measured effective potential $\hat{V}$ of the colloids as long as it occurs on a 
length scale which is comparable to or smaller than the typical distance $\ell_{\rm msd}$ along the $y$-axis which each particle 
explores during the acquisition of the images by the camera.
The images aquired during the experiments allow one to estimate such a mean-square 
displacement $\ell_{\rm msd}$ to be of the order of tens of $\mu$m \cite{vogt:2009a,troendle:2009,soyka:2008}.
The projection along the $y$-axis, which yields the density $\hat{\rho}(x)$, effectively causes a broadening of 
$\hat{\rho}(x)$  compared to the case of straight (ideal) chemical steps with $x_s(y)=\textit{const}$.
In addition, locally a smooth intrinsic chemical gradient of the step leads to such an effect, too.
For illustration purposes, we first consider  a single chemical step, which is ideally located at $x=0$ and 
which generates a potential $\Phi(x,z,T)$. 
(This reasoning can be extended to the periodic chemical pattern we are presently interested in  by assuming additivity of the 
forces, see Ref.~\cite{troendle:2010} for details.)
In order to estimate the effect of these variations of the actual position of the step along the $y$-axis, we assume that
the local position $x_s(y)$ of the step does not change significantly along the $y$-axis on the scale of the radius $R$ of 
the colloid, so that on this scale it can be considered as ideal and therefore  generates a potential $\Phi(x-x_s(y),z,T)$. 
For $x_s(y)$ we assume an effective Gaussian distribution $p(x_s)$ along the $y$-axis, with zero average and standard 
deviation $\Delta x$.
Accordingly, the projection $l^{-1}\int_0^l\upd y$ along the $y$-axis turns into
$\int_{-\infty}^{\infty}\upd x_s p(x_s)$ and affects the resulting projected density
$\hat{\rho}(x)$ and the resulting potential $\hat{V}(x)$ [see \eref{eq:vhat}].
\subsection{Particle-substrate distance}
The theoretical knowledge of the total potential $\Phi(x,z,T)$ [see \eref{eq:phitot}] allows one to predict the 
particle-substrate distance  $z$ as a function of the lateral variable $x$ and  temperature, a quantity which 
is not  accessible to the experiments discussed below.
As anticipated above, the rather small value of $G$ in \eref{eq:phig} is responsible for rather large fluctuations 
of the particle-substrate distance $z$ around the position $z=z_{\rm min}(x)$ at which the potential $\Phi(x,z,T)$ has a 
minimum as a function of $z$ for a fixed lateral position $x$ of the particle and which corresponds to the position 
of mechanical equilibrium.
In the presence of these large fluctuations it is convenient to consider the $x$-dependent  
mean particle-substrate distance $z_{\rm avg}(x)$, which is determined by the probability distribution function $\bar{P}$ 
of the colloid, i.e., by the total potential as\footnote[2]{%
Note that the Derjaguin approximation holds only for distances which are small on the scale of the particle size
(a detailed analysis of its applicability for the system under consideration is given in Ref.~\cite{troendle:2010}). 
However, in \eref{eq:avg} also large values of $z$ occur.
But at these large particle-wall distances the critical Casimir force as well as the electrostatic force are
negligibly small compared to the gravitational force, so that using this approximation is 
nonetheless not detrimental.
In principle, the integration in \eref{eq:avg} is limited by the vertical extension of the 
experimental sample cell of around $200\mu$m.
However, due to the gravitational contribution to the potential, de facto no colloidal particle moves out 
of the vertical field of view of the imaging objective.
Thus the integration in \eref{eq:avg} can be taken to run up to infinity without quantitatively relevant
consequences because contributions from large $z$ are strongly suppressed.
}
\begin{equation}
  \label{eq:avg}
  z_{\rm avg}(x)=\frac{1}{N(x)}\int_0^\infty\upd\,z\;z\;\exp\{-\Phi(x,z,T)/(k_BT)\},
\end{equation}
where $N(x)=\int_0^\infty\upd\,z\;\exp\{-\Phi(x,z,T)/(k_BT)\}$ is the normalization.
In order to describe the thermal fluctuations of the vertical position of the colloid it is convenient to consider 
the probability $\bar{P}_<(z;x)$ that for a fixed lateral position $x$ the colloid has a surface-to-surface distance 
from the substrate smaller than a given $z$:
\begin{equation}
  \label{eq:probability}
  \bar{P}_<(z;x)=\frac{1}{N(x)}\int_0^z\upd\,z'\;\exp\{-\Phi(x,z',T)/(k_BT)\}.
\end{equation}
In order to generalize the notion of ``standard deviation'' to the present case of an asymmetric distribution of the 
particle-substrate  distances at fixed lateral position, we define a lower value $z_{\rm low}(x)$ and an upper value 
$z_{\rm upp}(x)$ of the particle-substrate distances such that
\begin{equation}
  \label{eq:lowupp}
  \bar{P}_<(z_{\rm low}(x);x)=0.159\qquad \mbox{and}\qquad \bar{P}_<(z_{\rm upp}(x);x)=1-0.159,
\end{equation}
so that the probability of the colloid to be at a distance $z$ with $z_{\rm low}(x)<z<z_{\rm upp}(x)$
is $\simeq68\%$, whereas the probability to find it at distances smaller (larger) than $z_{\rm low}(x)$
($z_{\rm upp}(x)$) is $\simeq16\%$; these two properties  define the standard deviation for a Gaussian distribution.
\par
\begin{figure}
  \hspace*{-20mm}
  \begin{minipage}{170mm}
    \subfigure[$x=0$]%
    {\includegraphics{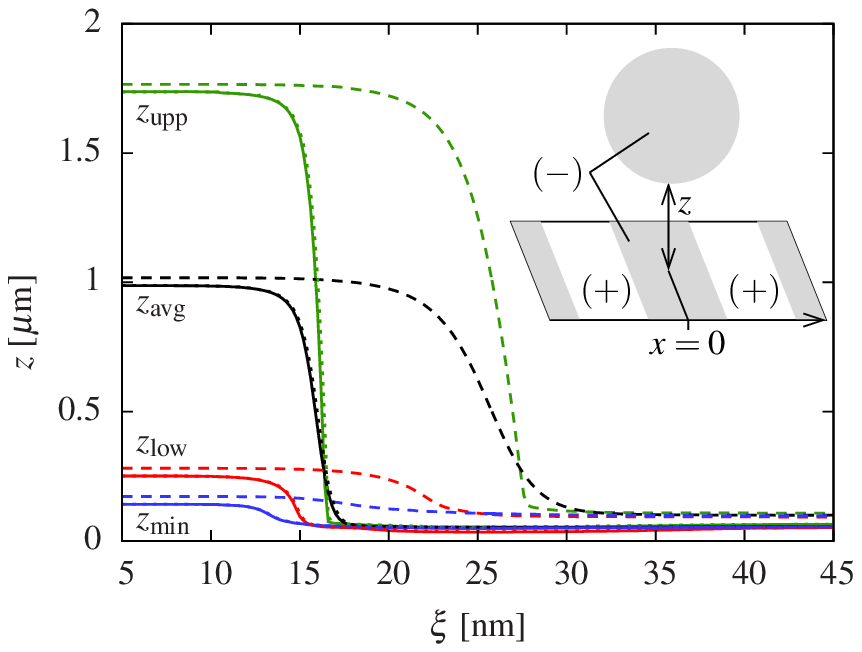}}
    \subfigure[$x=P/2$]%
    {\includegraphics{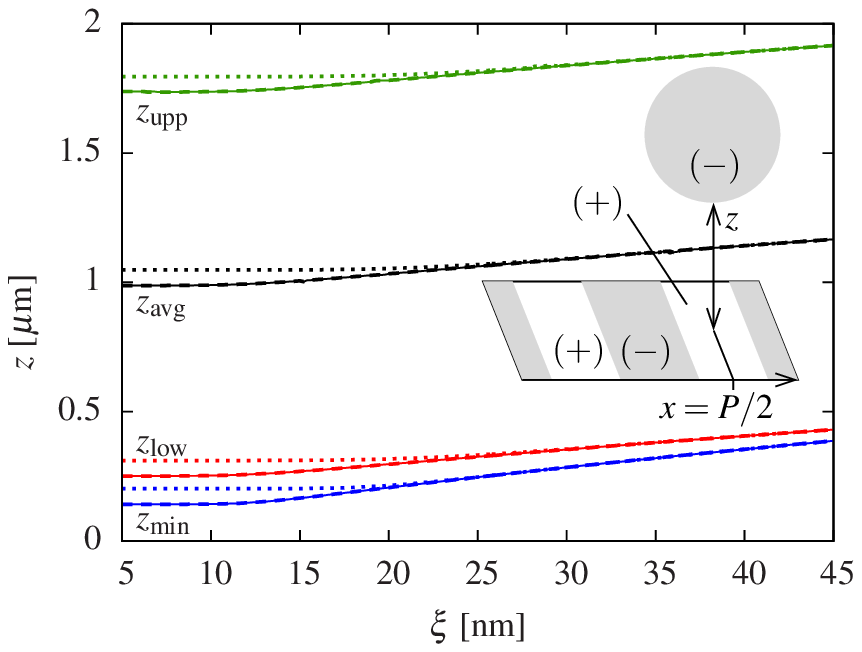}}
  \end{minipage}
  \caption{%
    Distances of the colloid from the substrate, for a fixed lateral position (a) $x=0$ and (b) $x=P/2$, as a function
    of the bulk correlation length $\xi$.
    Here, $P=1.8\mu$m, $R=1.2\mu$m, $L_-=0.9\mu$m, whereas $z_{\rm upp}$, $z_{\rm avg}$, $z_{\rm low}$, and $z_{\rm min}$
    indicate the upper, average, lower, and potential-minimum distances of the colloid, respectively (see the main text).
    The solid lines correspond to the choice $z_0^-=z_0^+=0.09\mu$m, the dashed lines to $z_0^-=0.12\mu$m and $z_0^+=0.09\mu$m,
    and the dotted lines to $z_0^-=0.09\mu$m and $z_0^+=0.15\mu$m.
    As expected, the fluctuations of the particle-substrate distance $z$ for a colloid opposite to an attractive stripe
    [$x=0$, panel (a)] decrease significantly upon increasing the correlation length $\xi$, due to the fact that the
    particle is mainly localized around the deep potential well which forms as a consequence of the  action of the 
    critical Casimir force.
    In fact, $z_{\rm upp, avg, low, min}\simeq z_0^-$ for large $\xi$.
    The actual value of $z_0^-$ strongly affects the behavior of the particle, both in determining the values
    of $z_{\rm upp, avg, low, min}$ close to the critical point and in setting the threshold value $\xi^*$ of $\xi$ at 
    which one observes such a sharp transition towards strong spatial confinement for $\xi>\xi^*$. 
    From panel (a) one infers, e.g., that $\xi^*\simeq 17$nm and $\xi^*\simeq 27$nm for the solid and dashed curves, 
    respectively.
    On the other hand, the dependence on $z_0^+$ is not pronounced and indeed in panel (a) the dotted lines practically coincide
    with the solid ones.
    For a colloid opposite to a repulsive stripe [$x=P/2$, panel (b)], due to the repulsive nature of both the
    electrostatic and the critical Casimir force and the very weak gravitational attraction one has
    $z_{\rm min}<z_{\rm low}$ and the average position is typically of the order of $1\mu$m with fluctuations
    of the same order.
    As anticipated, also in this case the actual value of $z_0^+$ is not very relevant, in particular for large values of $\xi$.
    Analogously, the behavior of the colloid at $x=P/2$ is not affected by the choice of $z_0^-$ and in panel (b) the
    dashed lines practically coincide with the solid ones.
    }
  \label{fig:dista}
\end{figure}
\par
Figure \ref{fig:dista} shows the behavior of the colloid with $(-)$ BC above a chemically patterned substrate for
the two lateral positions (a) $x=0$ and (b) $x=P/2$ at which the colloid is floating above the center of a $(-)$ 
and of a $(+)$ stripe, respectively, as a function of the correlation length $\xi$.
In \fref{fig:dista} the values of the geometrical parameters ($P$, $R$, $L_-$) are chosen to correspond to the actual 
experimental conditions, whereas the  parameters $z_0^\pm$ governing the electrostatic repulsion from the substrate 
are varied within a range which was determined by previous independent experiments.
Panel (a) clearly demonstrates that the fluctuations of $z$ for a colloid opposite to an attractive stripe decrease 
significantly upon increasing the correlation length $\xi$ above a certain threshold value $\xi^*$ which
depends on $z_0^-$.
This reflects the emergence of the deep potential well shown in \fref{fig:3d} which results from the competition 
between an increasingly attractive critical Casimir force and a repulsive electrostatic repulsion, the former being 
always overwhelmed by the latter around $z=z_0^-$.
Indeed, for small values of $\xi$ the average position $z_{\rm avg}(x=0)$ is typically determined by the competition 
between the electrostatic repulsion and the gravitational part, such that $z_{\rm avg}(x=0)\simeq z_0^- + k_BT/G
\simeq1\mu$m with fluctuations of the order of $k_BT/G \simeq1\mu$m.
On the other hand, for larger values of $\xi$, one has $z_{\rm upp, avg, low, min}(x=0)\simeq z_0^-$ and there are only 
very small thermal fluctuations of the particle-substrate distance, at most  a few tens of nm.
Depending on the relative strength of the electrostatic repulsion and the critical Casimir attraction,
one can have $z_{\rm min}<z_{\rm low}$ if the former dominates the latter, i.e., at small values of $\xi$ or the 
opposite at large  values of $\xi$.
From \fref{fig:dista}(a) one concludes that the choice of $z_0^-$ strongly affects the behavior of the particle,
both in determining the values of $z_{\rm upp, avg, low, min}$ and in setting the threshold value $\xi^*$ of $\xi$ above
which the particle becomes strongly confined, whereas the dependence on $z_0^+$ is negligible for the behavior at
$x=0$, because $z_0^+$ controls the electrostatic interaction with the adjacent stripe.
Analogously, the behavior of a colloid at $x=P/2$ as shown in \fref{fig:dista}(b), i.e., opposite to a repulsive stripe 
is not affected by the choice of $z_0^-$.
However, for this configuration, also the actual value of $z_0^+$ does not affect significantly the resulting behavior 
of the particle-substrate distance at $x=P/2$, in particular for large values of $\xi$.
Indeed, due to the repulsive nature of both the electrostatic and the critical Casimir force and the weak gravitational
attraction, the average position is typically of the order of $z_{\rm avg}(x=P/2)\simeq z_0^+ + k_BT/G\simeq k_BT/G$  
(with an additional linear contribution $\propto \xi$ for large values of $\xi$) allowing fluctuations of the order 
of $k_BT/G\simeq1\mu$m.
\par
\begin{figure}
  \hspace*{20mm}
    \includegraphics{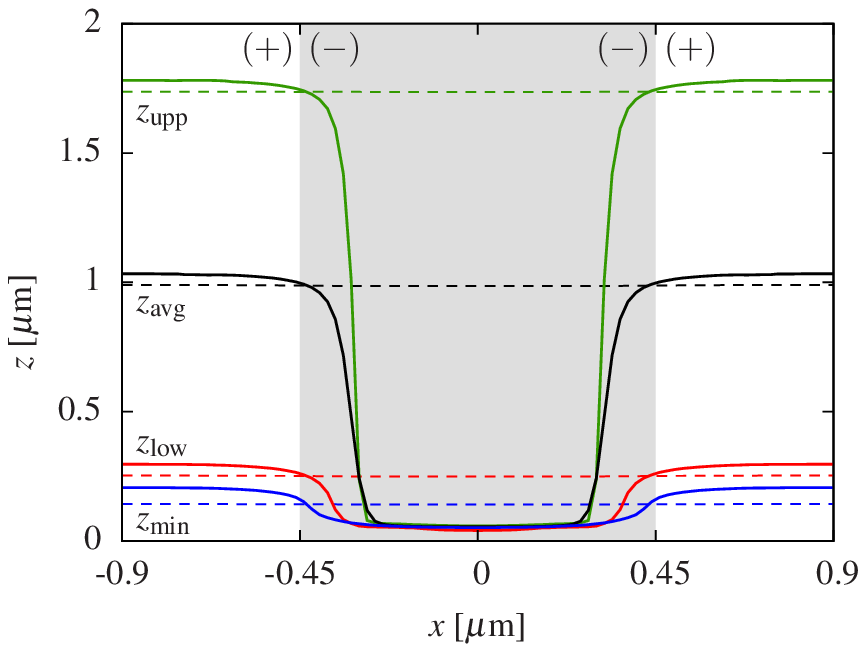}
  \caption{%
            Particle-substrate distance $z$ (characterized via $z_{\rm upp, avg, low}$, see the main text) and position 
            of mechanical equilibrium $z_{\rm min}$ as functions of the lateral coordinate of the particle for $\xi=10$nm 
            (dashed curves) and $\xi=20$nm (solid curves).
            Here, $z_0^-=z_0^+=0.09\mu$m, $P=1.8\mu$m, $R=1.2\mu$m, and $L_-=0.9\mu$m.
            For small values of the correlation length $\xi$ the various characteristic distances are almost independent of the 
            lateral coordinate, whereas above a certain threshold value $\xi^*$ of $\xi$ (see also \fref{fig:dista})
            the colloid opposite to the attractive stripe is strongly confined close to the wall at a distance
            $z\simeq z_0^-$ with almost no fluctuations.
          }
  \label{fig:dist}
\end{figure}
In \fref{fig:dist} the particle-substrate distance (characterized via $z_{\rm avg,upp,low}$) and the position of mechanical 
equilibrium $z_{\rm min}$ are reported as functions of the lateral coordinate of the colloid within the period $P$ and for 
two different temperatures, i.e., two different values of $\xi$.
In \fref{fig:dist}, the choice of the parameters corresponds to actually experimentally accessible values.
Figure~\ref{fig:dist} clearly shows that for $\xi=10$nm (dashed lines) the particle-substrate distance is
laterally constant and the colloid does not react to the presence of the chemical pattern on the substrate, apart 
for a possible effect due to a change in the electrostatic interaction (which here is taken to be the same 
on the different stripes). 
On the other hand, for $\xi=20$nm (solid line) the colloid is strongly attracted to that part of the substrate 
with the same preferential adsorption (in \fref{fig:dist} indicated by a shaded background), as a consequence of the emerging 
critical Casimir forces. 
Indeed for $|x|\lesssim L_-/2$ the particle is abruptly localized at a distance $z\simeq z_0^-$, which is primarily 
set by the electrostatic repulsion and which corresponds to the position $z_{\rm min}(0)$ of the minimum of the potential, 
with almost no thermal fluctuations.
In the region above the repulsive stripes (in \fref{fig:dist} indicated by a white background) and within the 
temperature range considered here, instead,
the repulsive critical Casimir force pushes the colloid only slightly further away from the substrate, by a distance of the order
of the increasing correlation length $\xi$. 
But the amplitude of the thermal fluctuations of the particle-substrate distance 
are barely affected by the onset of the repulsive critical Casimir force.
\par
The analysis of the particle-substrate distance shows that the behavior of the colloidal particle and
the resulting potential are drastically influenced by the strong attraction of the colloid close to an attractive stripe.
Accordingly, the electrostatic repulsion from this stripe, i.e., the value of $z_0^-$, affects significantly the total
potential and has therefore to be considered carefully in the comparison between theoretical predictions and the
experimental data.
On the contrary, the actual value of $z_0^+$ as well as the actual value of $\kappa$ do not significantly affect the 
resulting behavior of the potentials as long as they are within the range appropriate for the present experiment 
which can be inferred from previous, independent measurements \cite{gambassi:2009}.
%
\section{Experimental setup\label{sec:exp}}
In the present context of critical Casimir forces, surfaces with a periodic pattern of alternating stripes 
with opposite adsorption preferences for the two components of the 
binary mixture of water and lutidine were obtained by micro-contact printing ($\mu$CP) of alkane\-thiols. 
Since there is a vast literature covering this technique \cite{kumar:1994,Menard:2004,Qin:2010} here we shall describe 
it only briefly. 
Stamps were prepared by casting poly(dimethylsiloxane) (PMDS) onto a master which was topographically structured by a 
lithographic process. 
After the PMDS is cured one obtains an elastic stamp which exhibits the negative structure of the master. 
In the next step the stamp is wetted with a 1mM ethanolic solution of nonpolar $\mathrm{HS(CH_2)_{17}-CH_3}$ thiol.  
When the excess liquid is removed, the stamp is brought into mechanical contact for several seconds with a 
glass substrate coated by a $30$nm layer of gold. 
This results in hydrophobic regions the geometry of which corresponds to the protruding parts of the stamp. 
Finally, the substrate is dipped into a 1mM ethanolic solution of polar $\mathrm{HS(CH_2)_{11}-OH}$ thiol for several minutes 
in order to render the remaining bare gold surface hydrophilic. 
With this technique we were able to produce periodic arrays of hydrophilic and hydrophobic stripes of widths between 
$0.9\mu$m and $3\mu$m over a typical total extension of $0.5$cm$^2$. 
The typical lateral edge resolution of the chemical structures is of the order of $50-100$nm. 
In previous studies~\cite{soyka:2008} the chemical patterns were created by removing locally a hydrophobic monolayer from a 
hydrophilic surfaces with a focused ion beam (FIB). 
Although under ideal conditions the resolution of FIB is within the range of several nanometers, in the case of 
non-conducting surfaces we observed distortions/deflections of the ion beam due to an electrostatic charging of the glass surface, 
leading to deviations of the chemical steps from straight lines. 
This causes considerable discrepancies between the measured critical Casimir forces acting on particles near such 
surfaces and the corresponding theoretical predictions, which assume straight and sharp boundaries for the stripes
\cite{troendle:2009}. 
Since such charging effects are basically absent in $\mu$CP, here we have an improved control of the geometrical structure 
of the imprinted chemical pattern and therefore much better agreement with theory (see below).
\par
In order to probe the critical Casimir forces which act on a colloidal particle exposed to the patterned substrates prepared 
as described above, we used polystyrene (PS) probe particles with radius $R=1.2\mu$m. 
Their surface charge, as provided from the manufacturer, is $10\mu$C/cm$^2$ which renders them hydrophilic, 
realizing $(-)$ boundary conditions. 
The particles were dissolved in a critical water-2,6-lutidine mixture which has a lower critical demixing point at a 
lutidine mass fraction of $c_L^c \cong 0.286$ and a critical temperature of $T_c=307$K. 
Particle positions were monitored by digital video microscopy which allows one to track the projection of their centers onto 
the substrate plane with a spatial resolution of about 50 nm. 
For each measurement the particle trajectories were recorded for 40 minutes with a frequency of 4 frames per second.
Within the field of view of $290\mu$m$\times225\mu$m around 20 to 40 particles have been monitored, corresponding to a projected
density of one particle per square of the size of 25$\times$25 to 17$\times$17 particle diameters.
Thus, the particle density was sufficiently small to exclude the presence of critical Casimir forces among neighboring colloids, such that only their interaction with the patterned substrate 
is probed during the measurements. 
\par
The temperature of the sample was controlled by the procedure described in detail in Ref.~\cite{soyka:2008}. 
With this setup, temperatures close to $T_c$ could be stabilized within $10\ \mathrm{mK}$ over several hours. 
In contrast to temperature \emph{changes}, which could be resolved within mK accuracy, larger errors in the 
determination of \emph{absolute} temperatures (in particular the measured critical temperature) occur. 
This is due to the fact, that the metal-resistance thermometer (Pt100) could not be placed within the sample 
cell but it was instead attached outside  but close to the field of view. 
Indeed, the thermometer measures a temperature $T^{\rm out}$ lower than the actual temperature $T$ inside
the sample cell.
We associate the temperature $T^{\rm out}=T_c^{\rm out}$ with the critical temperature $T=T_c^{\rm exp}$ of the solvent 
in the sample at that particular temperature for which critical opalescence is observed when shining a laser beam
into the sample cell.
This leads to systematic errors on the absolute temperature of $T_c^{\rm out}$ and $T_c^{\rm exp}$ of the order of 
$50\ \mathrm{mK}$.
Using the assumption that $T-T^{\rm out}=\textit{const}$ (this constant could not be determined and
is inter alia dependent on the ambient temperature), we have measured the temperature difference $\Delta T=T_c^{\rm out}-T^{\rm out}$
and identified it with $T_c^{\rm exp}-T\equiv\Delta T$.
Since the comparison with the theoretical predictions depends crucially on the actual value $T_c\simeq307$K
of the critical temperature, in the analysis below we account for such a possible systematic error 
by considering $\Delta T_c^*=T_c-T_c^{\rm exp}$ as an additional fitting parameter,
where $T_c\simeq307$K is the actual critical temperature of the water-lutidine mixture.
\section{Comparison of theory with experiments}
%
The total potential $\Phi$ of the forces [\eref{eq:phitot}] acting on the colloidal particle has been calculated on the basis of 
the Derjaguin approximation with $k_{(\pm,-)}$ [see Eqs.~\eqref{eq:planar-force} and \eqref{eq:step-omega-da}] obtained 
from Monte Carlo simulations. 
In the following we shall use for the scaling functions $k_{(\pm,-)}$ of the critical Casimir force between two planar 
walls with $(\pm,-)$ BC the numerical estimate referred to as ``approximation (\textit{i})'' in Figs.~9 and 10 of 
Ref.~\cite{vasilyev:2009}. 
We have checked that different~\cite{vasilyev:2009} or more recent and accurate~\cite{hasenbusch:2010} estimates for 
$k_{(\pm,-)}$  actually lead to essentially the same  effective potentials, the only difference being  a small 
additional overall shift of the resulting fitted value of the critical temperature (see further below).
A detailed analysis in spatial dimension $d=4$ suggests that the Derjaguin approximation we have used in our theoretical 
predictions for $d=3$ should be rather accurate in describing the actual potential of the colloid within the range of 
parameters experimentally studied here \cite{troendle:2010}.
As anticipated above, in order to predict the effective potential $\hat{V}$ we need to fix also the value of the 
parameters which determine the electrostatic interaction. 
For the comparison between theory and experiment we fix the screening length $\kappa^{-1}$ to the value $\kappa^{-1}=12$nm 
which has been reported from independent measurements on the same system (see, e.g., Ref.~\cite{gambassi:2009}).
In order to fit our theoretical predictions to the experimental data, we vary instead the unknown values of the 
parameters $z_0^\pm$ [Eqs.~\eqref{eq:elhomog} and \eqref{eq:phiel}] within the 
range $0.08-0.15\mu$m, which can be reasonably expected on the basis of previous measurements on homogeneous 
substrates \cite{gambassi:2009}.
However, the results for $\delta\hat{V}(0)$ are hardly affected by the particular choice of $z_0^+$ so that in the following 
we keep it fixed at $z_0^+=0.09\mu$m.
The amplitude $\xi_0^+$ of the correlation length has been determined by independent experiments
as $\xi_0^+\simeq0.20\pm0.02$nm (see, e.g., Tab.~III in Ref.~\cite{gambassi:2009}) so that we vary 
$\xi_0^+$ within the range $0.18-0.22$nm in order to obtain the best fit to the experimental
data by the theoretical scaling functions.
Moreover, as mentioned above, the experimental uncertainty in the determination of the absolute value of
the critical temperature $T_c^{\rm exp}$ is taken into account 
by considering as an additional fitting parameter the shift $\Delta T_c^*=T_c-T_c^{\rm exp}$ (see the end of Sec.~\ref{sec:exp})
of up to $|\Delta T_c^*|\simeq100$mK. 
(The values of $\Delta T_c^*$ may be different for the individual stripe widths $L_-$
which characterize the substrates investigated in
independent experimental runs.)
However, the \emph{relative} uncertainty in the determination of the temperature 
within a single experimental run for a given stripe width $L_-$ is 
smaller than $\pm10$mK (see above).
Thus, for the comparison carried out below, we are left with $\Delta T_c^*$, $z_0^-$, $\xi_0^+$, and $\Delta x$  
(see Subsec.~\ref{sec:non-ideal})
as fitting parameters which, however, are all limited to rather small ranges of values in order to be in accordance 
with independent and previous experimental results.
%

\subsection{Depth of the potential}
\begin{figure} 
  \hspace*{-10mm}
  \begin{minipage}{160mm}
  \caption{%
   Depth $\Delta\hat{V}$ of the effective potentials of a colloid close to  patterned substrates, 
   with periods $P=L_-+L_+$ and    $(-)$ stripe widths $L_-$ given in Tab.~\ref{tab:best}, as a function of the 
   temperature deviation from the critical temperature.
   The symbols represent the experimental data which are affected by an experimental uncertainty of
   $\pm0.15k_BT$ for the potential depth and of $\pm0.01$K for the temperature differences between
   the various data points belonging to one and the same value for $L_-$.
  }   
  \label{fig:depth}
    \subfigure[%
   $\Delta\hat{V}$ as a function of $\Delta T = T_c^{\rm exp}-T$.
   (For better visibility, the data for $L_-=3.25\mu$m are shifted by $0.1$K along the $\Delta T$ axis, as indicated by the 
   red arrow.)
   The solid lines represent the best fit of the theory to the experimental data for each \emph{individual} value of $L_-$
   with fixed parameters $\kappa$ and $z_0^+$ which do not affect the resulting behavior significantly as discussed in the main text. 
   $\xi_0^+$ has been varied within the reasonable range $0.18$nm$\lesssim\xi_0^+\lesssim0.22$nm known from the literature.
   The values of all parameters corresponding to the solid lines are given in Tab.~\ref{tab:best}.
   Clearly, for each individual $L_-$ the actual critical temperature $T_c$ is shifted by $\Delta T_c^*=T_c-T_c^{\rm exp}$ 
   from the value $T_c^{\rm exp}$ determined experimentally, with $\Delta T_c^*$ being within the experimental accuracy 
   $|\Delta T_c^*|\lesssim0.1$K.
   The dashed curves represent a \emph{common} fit to all data.
   For this latter fit we have assumed $\xi_0^+$ and $z_0^-$ to be the same for the various independent experimental runs,
   independently of the corresponding value of $L_-$, whereas $\Delta T_c^*=T_c-T_c^{\rm exp}$ (besides $\Delta x$ as determined below)
  is used for adjustment to the data for each individual $L_-$.
   The best fit for all these parameters renders $\xi_0^+=0.22$nm, $z_0^-=0.11\mu$m,
    and the values reported in Tab.~\ref{tab:common}.
    ]{\includegraphics{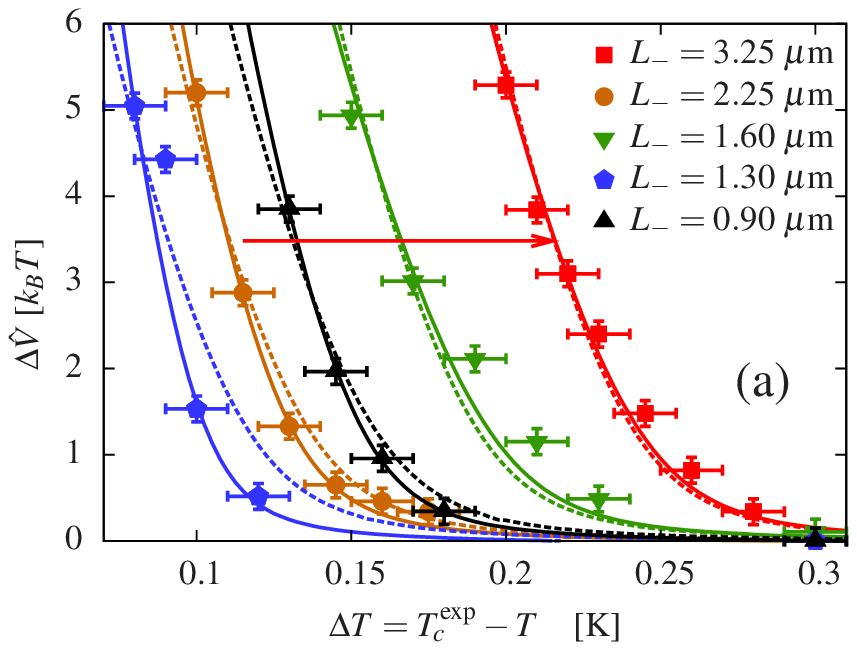}}
    \hfill
    \subfigure[%
      $\Delta\hat{V}$ as a function of $\Delta T + \Delta T_c^*= T_c-T$ using the values of $\Delta T_c^*$ for the
      best common fit (see dashed lines in (a)) given in Tab.~\ref{tab:common}.
      This plot takes into account the experimental uncertainty of measuring the absolute value of $T_c^{\rm exp}$.
      The data points, which have been shifted accordingly, basically fall on top
      of each other within their error bars.
      The dashed lines correspond to the dashed ones shown in (a) and represent the theoretical predictions.
      For $L_-\ge1.60\mu$m the various curves are almost indistinguishable from each other, 
      whereas for $L_-\le1.30\mu$m the 
      potential depth is effectively reduced due to the interference of the effects of two neighboring
      steps and due to the fact that the steps are non-ideal.
    ]{\includegraphics{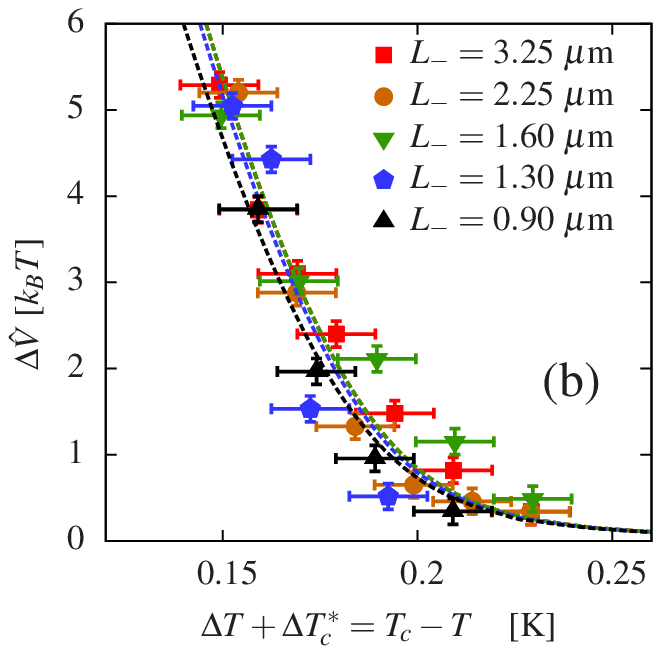}}
  \end{minipage}
\end{figure}
The depth of the measured laterally confining potential is given by 
$\Delta\hat{V}=-\delta\hat{V}(0)>0$ [see \eref{eq:vhat}].
Figure~\ref{fig:depth}(a) shows $\Delta\hat{V}$ as a function of the deviation
$\Delta T=T_c^{\rm exp}-T$ of the temperature from the experimentally determined critical demixing temperature $T_c^{\rm exp}$,
for various stripe widths $L_-$ and measured in independent experimental runs as described above.
The periodicities $P$ of the various patterns  the stripes belong to are determined from the photolithography mask
and are given in Tab.~\ref{tab:best}.
The potential depth $\Delta\hat{V}$ is determined as the difference of the potential between its value at
the center of a repulsive stripe and at the center of an attractive stripe, carrying an uncertainty
of $\pm0.15k_BT$ near the critical temperature $T_c\simeq307$K.
Upon approaching $T_c$, for a stripe width $L_-\lesssim1.5\mu$m the effects of two adjacent chemical steps interfere,
which results in an effectively reduced potential depth compared to the case of a very wide stripe (see
the dashed curves in \fref{fig:depth}(b); in \fref{fig:depth}(a) the experimental uncertainty in determining the critical
temperature $T_c^{\rm exp}$ is responsible for the relative displacements among the various curves).

The effect of non-ideal chemical steps may result in an effectively reduced potential depth as well
(see, for example, \fref{fig:depth}(b) and, c.f., \fref{fig:narrow}(b)).
Accordingly, in comparing the  experimental data for $\Delta\hat{V}$ with the theoretical predictions we allow for an uncertainty 
in the local position of the boundary between adjacent stripes, as described in Subsec.~\ref{sec:non-ideal}.
However, as will be shown below, we are able to determine $\Delta x$ rather precisely from the lateral
variation of the potential.
The values of $\Delta x$, which yield the best agreement and which are used for the comparison shown in
\fref{fig:depth}, are given in Tab.~\ref{tab:best}.
%
%
On the basis of our theoretical analysis, it turns out that the temperature dependence of the potential depth 
becomes independent of the stripe width when the latter is sufficiently large.
Adopting for the geometrical and physical parameters the values corresponding to the present experimental conditions, 
this is expected to occur for $L_-\gtrsim2\mu$m.
From \fref{fig:depth} one can see, however, that the experimentally determined data for individual experimental runs
are shifted along the temperature axis with respect to each other, which reflects the uncertainty of up to 100mK 
in measuring the absolute value of the  critical temperature.
On the other hand, within a single individual experimental run corresponding to a certain stripe width, 
the temperature difference between 
the various data points can be measured with the high accuracy of less than $10$mK.
Upon comparing the experimental data shown as symbols in \fref{fig:depth} with our theoretical predictions
we take this into account by introducing, for each value of $L_-$, a possible shift 
$\Delta T_c^*(L_-)=T_c-T_c^{\rm exp}(L_-)$ between the actual critical temperature $T_c$ and the value $T_c^{\rm exp}$ determined 
in that particular experiment. 
Accordingly, for the data corresponding to a certain $L_-$, the actual distance from the critical point is 
given by $T_c - T = \Delta T + \Delta T_c^*(L_-)$.
\par
The dashed curves in \fref{fig:depth} correspond to the \emph{common} fit to all experimental data  which leads to
the fitting parameters  $\xi_0^+$ and $z_0^-$ taking the same values $\xi_0^+=0.22$nm and $z_0^-=0.11\mu$m for all
$L_-$ as obtained from the least-square method.
On the other hand, the individual temperature shifts $\Delta T_c^*(L_-)$ are given in Tab.~\ref{tab:common}
for each data set corresponding to a single value of $L_-$.
The values of $\xi_0^+$ and $z_0^-$  are both in agreement with independent previous findings \cite{gambassi:2009}.
Figure~\ref{fig:depth}(b) shows the depth of the potentials as a function of $\Delta T+\Delta T_c^*(L_-)=T_c-T$,
i.e.,  shifted along the temperature axis by $\Delta T_c^*(L_-)$ as given in Tab.~\ref{tab:common}.
This accounts for the uncertainty in determining experimentally the absolute value of the temperature
(see the discussion at the end of Sec.~\ref{sec:exp}), and indeed,
as expected for significantly wide stripes $L_-\gg\xi$, the data overlap with each other within the 
error bars, reflecting that in this limit $\Delta\hat{V}$ is independent of $L_-$.
%
\par
The solid lines in \fref{fig:depth}(a) correspond to fits of the theoretical predictions to each \emph{individual}
experiment dealing with a specific stripe width.
Distinct from the previous common fit, here for each stripe width $L_-$ we allow for a variation of the values of 
$\xi_0^+$ and $z_0^-$ in addition to $\Delta T_c$ and $\Delta x$.
On the basis of a least-square fit, best agreement is obtained for the values given in Tab.~\ref{tab:best}, 
which agree with those obtained in independent previous studies \cite{gambassi:2009} within the 
corresponding experimental accuracy.
\begin{table}
  \linespread{1.0}\selectfont
  \tbl{%
  Values of the parameters for which best agreement is obtained between theory and the data 
  for $\delta\hat{V}(x)$ for each individual experimental run, corresponding to a single value of 
  $L_-$  (see \fref{fig:depth}(a)).
  The values $\kappa^{-1}=12$nm and $z_0^+=0.09\mu$m are fixed because their choice does not affect significantly
  the resulting theoretical  curves in \fref{fig:depth}.
  }%
  {\begin{tabular}{@{}lcccccc} \toprule
    $P$[$\mu$m] $^{\rm a}$ & 
    $L_-^{\rm exp}$[$\mu$m] $^{\rm b}$ & 
    $L_-$[$\mu$m]  $^{\rm c}$  & $\xi_0^+$[nm] & 
    $z_0^-$[$\mu$m] & 
    $\Delta T_c^*$[mK] & 
    $\Delta x$[$\mu$m] $^{\rm d}$ \\
    \colrule
    6.0 & 3.0 &  3.25 & 0.22 & 0.103 &  86 & 0.15 \\
    5.4 & 2.7 &  2.25 & 0.21 & 0.128 & -5  & 0.10 \\
    4.2 & 2.1 &  1.60 & 0.22 & 0.095 &  88 & 0.22 \\
    3.6 & 1.8 &  1.30 & 0.19 & 0.140 & -18 & 0.19 \\
    1.8 & 0.9 &  0.90 & 0.20 & 0.121 & -28 & 0.09 \\
    \botrule
  \end{tabular}
  }
  \tabnote{$^{\rm a}$Most of the experimental data for the potential $\Delta\hat{V}$ discussed here are not
  influenced by the actual value of $P$ as long as $P\gtrsim2L_-$. Accordingly, we assume the periodicity to be the one
  determined by the inscribed photolithographic mask pattern reported here.}
  \tabnote{$^{\rm b}$Width of the stripes of the photolithographic mask pattern. Due to the $\mu$CP 
  stamping process we estimate the uncertainty of the actual width of the thiol stripes to be up to $\pm0.5\mu$m.}
  \tabnote{$^{\rm c}$Value of the stripe width for which best agreement between theory and experiment is obtained.}
  \tabnote{$^{\rm d}$Value of $\Delta x$ for which best agreement between theory and experiment is obtained (see Figs.~\ref{fig:broad}
  and \ref{fig:narrow}).}
  \label{tab:best}
\end{table}
\begin{table}
  \linespread{1.0}\selectfont
  \tbl{%
  Values of the parameters for which best agreement is obtained between theory and all experimental data together (see \fref{fig:depth}(b)),
  so that the values of $\xi_0^+$ and $z_0^-$ are the same, and $\Delta T_c^*$ (in addition to $\Delta x$ determined from
  the shape of the potentials; see below) is the only parameter allowed to vary for the various stripe widths $L_-$.
  As  in Tab.~\ref{tab:best}, the values of  $\kappa^{-1}=12$nm and $z_0^+=0.09\mu$m are fixed.
  (For a description of the parameter values $P$, $L_-^{\rm exp}$, and $L_-$ see the footnotes in Tab.~\ref{tab:best}.)
  }%
  {\begin{tabular}{@{}lcccccc}%
    \toprule
    $P$[$\mu$m] $^{\rm a}$ & 
    $L_-^{\rm exp}$[$\mu$m] & 
    $L_-$[$\mu$m] & 
    $\xi_0^+$[nm] & 
    $z_0^-$[$\mu$m] & 
    $\Delta T_c^*$[mK]  & 
    $\Delta x$[$\mu$m]\\
    \colrule
     6.0 & 3.0  &  3.25 & 0.22 & 0.110 & 49 & 0.15 \\
     5.4 & 2.7  &  2.25 & 0.22 & 0.110 & 54 & 0.10\\
     4.2 & 2.1  &  1.60 & 0.22 & 0.110 &  0 & 0.22 \\
     3.6 & 1.8  &  1.30 & 0.22 & 0.110 & 72 & 0.19 \\
     1.8 & 0.9  &  0.90 & 0.22 & 0.110 & 29 & 0.09 \\
     \botrule
  \end{tabular}
  }
  \label{tab:common}
\end{table}
\subsection{Shape of the potentials}
\begin{figure}
  \hspace*{-15mm}
  \begin{minipage}{175mm}
  \caption{%
  Lateral variation of the effective potential $\delta\hat{V}(x)$ (see \eref{eq:vhat}) 
  for a colloid opposite to a chemically
    patterned substrate and immersed in the water-lutidine binary liquid mixture at its critical concentration for
    various temperatures $T_c^{\rm exp}-\Delta T$. 
    Symbols indicate experimental data, whereas the lines are the corresponding theoretical
    predictions for ideal (dashed lines) and non-ideal (solid lines) stripe patterns.
    The parameters used to calculate the theoretical curves are given in Tab.~\ref{tab:best}.
    }
    \label{fig:broad}
    \subfigure[$L_-=3.25\mu$m.
    From top to bottom the critical point is approached and the measured temperature deviations $\Delta T$ are 
    0.3, 0.18, 0.15 (0.153), 0.145 (0.14), 0.13 (0.128), 0.12, 0.11 (0.113), and 0.10 K 
    with an uncertainty of up to $\pm0.01$K with respect to each other.
    If indicated, the values in brackets are corrected values (but compatible within the experimental inaccuracy)
    which have been used for evaluating the theoretical predictions.
    ]{%
    \resizebox*{85mm}{!}{
    \includegraphics{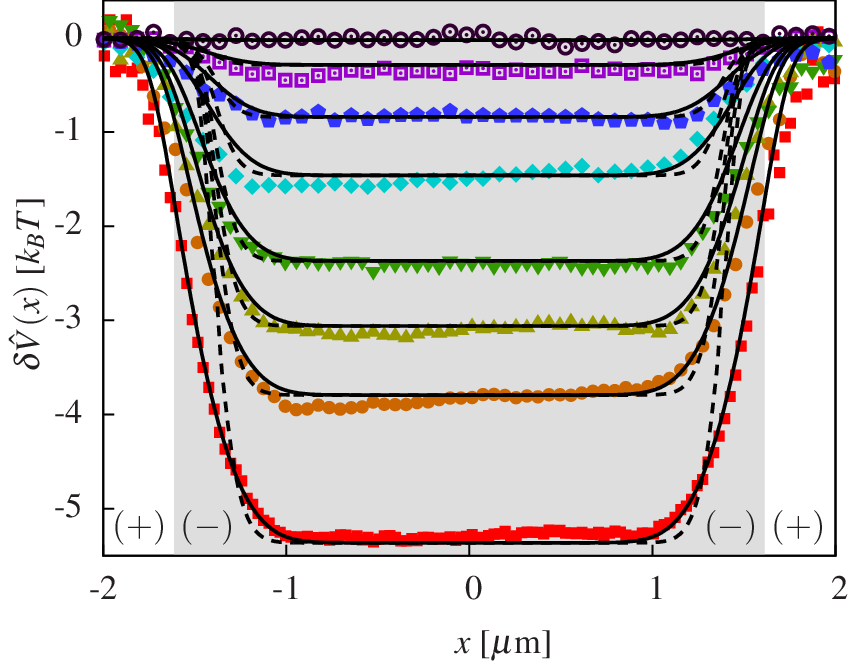}}}
    \hfill
    \subfigure[Same as (a) for $L_-=2.25\mu$m and $\Delta T=$  
    0.175 (0.165), 0.16 (0.152), 0.145 (0.143), 0.13, 0.115, 0.10 K.
    ]{%
    \resizebox*{85mm}{!}{
    \includegraphics{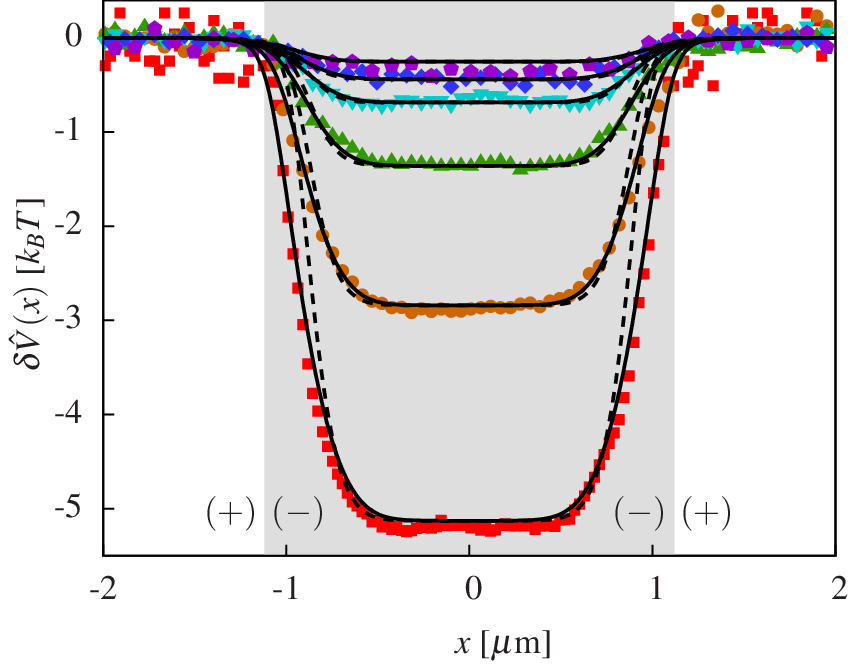}}}
  \end{minipage}
\end{figure}

\begin{figure}
  \hspace*{-15mm}
  \begin{minipage}{175mm}
  \caption{%
  Effective potential $\delta\hat{V}(x)$ (see \eref{eq:vhat}) for various temperatures $T_c^{\rm exp}-\Delta T$ 
  and relatively     small widths of the attractive stripes.
    Symbols indicate experimental data, whereas the lines are the corresponding theoretical
    predictions for ideal (dashed lines) and non-ideal (solid lines) stripe patterns.
    The parameters used to calculate the theoretical curves are given in Tab.~\ref{tab:best}.
    }
    \label{fig:narrow}
    \subfigure[Same as \fref{fig:broad}(a) for $L_-=1.60\mu$m and $\Delta T=$ 
    0.3, 0.23 (0.218), 0.21 (0.198), 0.19 (0.182), 0.17, 0.15 (0.153) K. 
    ]{%
    \resizebox*{85mm}{!}{
    \includegraphics{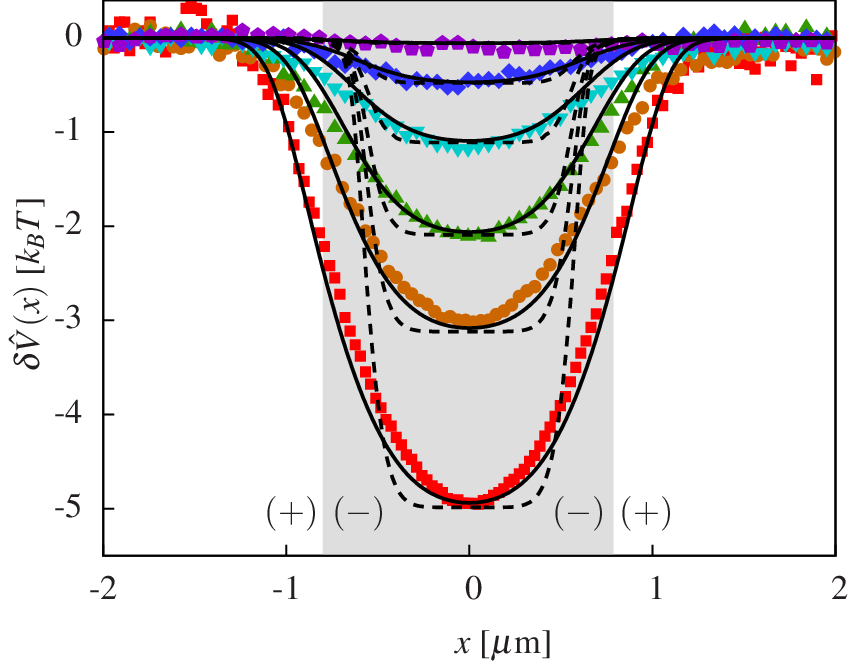}}}
    \hfill
    \subfigure[Same as (a) for $L_-=1.30\mu$m (left panel) and $L_-=0.9\mu$m (right panel) with
    $\Delta T=$ 0.3, 0.12, 0.1, 0.09 (0.083), 0.08 K for $L_-=1.30\mu$m 
    and $\Delta T=$ 0.3, 0.18, 0.16, 0.145, 0.13 K for $L_-=0.9\mu$m.
    ]{%
    \resizebox*{85mm}{!}{
  \includegraphics{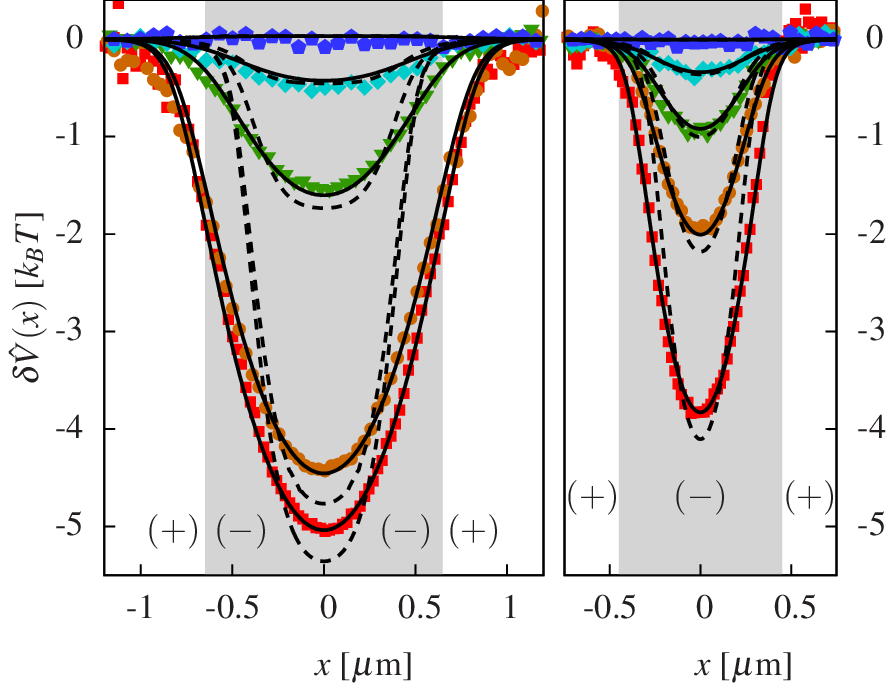}}}
  \end{minipage}
\end{figure}

%
Figures~\ref{fig:broad} and \ref{fig:narrow} show the total potential of the forces acting on the colloid 
as a function of its lateral position for various temperatures and for two stripe widths. 
Symbols represent the experimental data, whereas the solid and dashed lines are the 
corresponding theoretical predictions with $\xi_0^+$, $z_0^-$, and $\Delta T_c^*$  fixed to 
the values reported in Tab.~\ref{tab:best}, which have been determined from the fit of the depth $\Delta\hat{V}$
of the potential.
In addition, for a few cases and as indicated in the figure captions, we use the leeway provided by 
the experimental uncertainty of $\pm10$mK for the temperature value.
The dashed lines in Figs.~\ref{fig:broad} and \ref{fig:narrow} are based on the assumption that the stripe patterns 
are \emph{ideal} whereas the solid ones refer to \emph{non-ideal} patterns for which we fitted the parameters 
$\Delta x$ (see Subsec.~\ref{sec:non-ideal}) in order to obtain the best agreement between theoretical predictions 
and the set of experimental data at distinct temperatures.
The resulting values of $\Delta x$ for the various widths of the pattern  are reported in Tab.~\ref{tab:best}.
In fact, the broadness of the transition regions across the chemical steps of $\delta\hat{V}(x)$ between 
its extremal values  is affected practically exclusively by $\Delta x$ among the fitting parameters.
We have checked that the other parameters and potential additional effects such as 
polydispersity and weak adsorption cannot account for the discrepancies of the slopes of the dashed curves and
the experimentally determined ones (see also Ref.~\cite{troendle:2009}).
Thus, $\Delta x$ and $L_-$ are  determined de facto independently of the choice of $\xi_0^+$, $z_0^-$, 
and $\Delta T_c^*$ and in the following they can be  regarded fixed upon variation of the values of these latter parameters.
In particular, this is important for narrow stripes as shown in \fref{fig:narrow}, where even the depth of the 
potential depends on $L_-$ and $\Delta x$, in contrast to broader stripes.
\par
As anticipated, in the case of rather large stripe widths $L_-=3.25\mu$m and $L_-=2.25\mu$m
shown in \fref{fig:broad} and for the temperatures considered here the effects of 
two neighboring chemical steps do not interfere and the resulting
potential across the center of the attractive chemical stripe is flat.
Accordingly, from the comparison between the experimental data for the shapes of the potentials 
and the corresponding theoretical prediction we can infer the width $L_-$ of the stripes 
--- now treated as a fitting parameter --- independently of its initially assumed value 
$L_-^{\rm exp}$ determined by the width  of the photolithography mask.
Indeed, the actual width of the stripe imprinted by the mask may differ by up to $\pm0.5\mu$m 
from $L_-^{\rm exp}$ characterizing the mask itself.
It is only $L_-^{\rm exp}$  which is controlled during the preparation process \cite{vogt:2009a}.
On the other hand, the actual value of $P$ is the same as the one of the photolithography mask
because the periodicity is not affected by the stamping process.
(Note that, within the parameter range investigated in \fref{fig:broad}, the effects of neighboring chemical steps 
do not influence each other and therefore the actual value of the periodicity $P$ of the 
pattern does not affect the determination of $L_-$.)
In agreement with this observation, the comparison with the theoretical predictions leads to 
fitted values $L_-$ which are within the aforementioned range  indicated in Tab.~\ref{tab:best}.
From \fref{fig:broad} we find that the experimental techniques described above lead to rather sharp
chemical steps between two stripes with an uncertainty of only $\Delta x\le 0.15\mu$m. 
\par
The comparison for narrower stripe widths $L_-\lesssim2\mu$m is shown in \fref{fig:narrow}.
Even for these cases, in which the effects of two neighboring chemical steps do interfere, the experimental
data agree very well with the theoretical predictions obtained from the Derjaguin approximation.
Whereas for $L_-=1.6\mu$m and $L_-=1.3\mu$m we have to take into account an uncertainty $\Delta x$ of 
the position and associated with the shape of the chemical steps of about $0.2\mu$m, for the 
smallest stripe width $L_-=0.9\mu$m the chemical
pattern turns out to be almost ideal with $|\Delta x<0.1\mu$m (see Tab.~\ref{tab:best}).
\section{Conclusions and Outlook}
\subsection{Summary}
We have presented a detailed theoretical and experimental study of the effective solvent-mediated
forces acting on colloids suspended in a near-critical binary liquid mixture of water and 2,6-lutidine
and close to chemically patterned substrates.
In contrast to earlier investigations \cite{soyka:2008} we have obtained a chemical pattern of stripes
of different widths and with laterally alternating adsorption preferences for either lutidine [$(+)$ boundary condition]
or water [$(-)$ boundary condition] by monolayers of thiols deposited on gold-coated
substrates via microcontact printing.  
The solute colloidal particles --- $2.4\mu$m diameter polystyrene spheres --- were rendered
hydrophilic, corresponding to $(-)$ boundary conditions, due to their surface charge. 
Based on digital video microscopy our main experimental findings are the following.
\begin{itemize}
\item[1.] At the fixed critical composition of the water-lutidine mixture, upon raising the temperature $T$ towards
  its lower critical demixing point at $T=T_c$, \emph{lateral} forces acting on the colloidal particles arise
  gradually.
  These critical Casimir forces cause the colloidal particles to be attracted to those chemical stripes which have the same
  adsorption preference as the particles and to be repelled from those stripes with the opposite adsorption preference.
\item[2.] These lateral and normal contributions to the total effective forces (compare Subsec.~\ref{sec:allforces}) are 
  negligible at temperatures $T$ more than a few hundred mK away from the critical value $T_c$ but they increase 
  significantly upon approaching $T_c$. 
  Eventually, they lead to laterally confining potentials for the colloidal particles with potential depths of 
  several $k_BT$ (see \fref{fig:depth}).
  These potentials can be reversibly tuned by minute temperature changes.
\item[3.] Across a chemical step the critical Casimir potentials vary rather abruptly between two plateau values
  on a length scale of $\approx0.8\mu$m  (see \fref{fig:broad}).
  This indicates that the microscopic structures of the chemical steps formed by imprinted
  layers of alkanethiols with different endgroups are much sharper than those
  created previously by a focused ion beam \cite{soyka:2008}.
\item[4.] For rather  narrow  stripes of widths $\lesssim2\mu$m the effects of two neighboring
  chemical steps interfere and consequently reduce the potential depth and lead to rounded
  shapes of the effective potentials (see \fref{fig:narrow}).
\end{itemize}
These experimental observations can be consistently interpreted in terms of the occurrence of the
critical Casimir effect, and it is possible to quantitatively compare the measured potentials
with the corresponding theoretical predictions.
We have derived the effective potentials within the Derjaguin approximation (see Subsec.~\ref{sec:casimir} and
Ref.~\cite{troendle:2010}) based on universal scaling functions for the critical Casimir force between two plates with $(\pm,\pm)$ or
$(+,-)$ boundary conditions as obtained from Monte Carlo simulation data.
(Differences in estimates for these scaling functions as obtained from various Monte Carlo simulations
do not affect our results significantly.)
In contrast to the present experimental measurements we are able to theoretically analyze the spatially fully resolved
critical Casimir potential acting on a colloid.
The resulting energy landscape for a colloidal particle is rather rich due to the interplay of
several relevant forces, which we have taken into account (see \fref{fig:3d}).
Typically, thermal fluctuations lead to large fluctuations of the lateral position and the distance
between the colloidal particles and the substrate (see \fref{fig:dista}).
Upon approaching $T_c$, however, strong normal and lateral critical Casimir forces abruptly localize the
colloids very close to and above stripes exhibiting the same adsorption preference as the particles
(see \fref{fig:dist}).
In calculating the effectively one-dimensional, projected potentials for a single colloid, as they are obtained experimentally,
we also take into account the possibility that the chemical pattern is not ideal, in addition to other experimental uncertainties.
Not only the theoretical predictions for the potential depths agree with the experimental data (see \fref{fig:depth})
but also the correlation length $\xi$, as determined from our comparison, follows rather well the expected
universal power-law behavior and the associated non-universal amplitude $\xi_0$ is in agreement with previous
independent experimental determinations (see Tab.~\ref{tab:best}).
Moreover, the shapes of the potentials as a function of the lateral position of the colloid show good
agreement between the theory and the experiments (see Figs.~\ref{fig:broad} and \ref{fig:narrow}).
Since critical Casimir forces probe rather sensitively the details of the geometry of the patterns
\cite{troendle:2009}, we find that the chemical steps obtained by the experimental method used here are
sharp, with deviations of only a few percent in terms of the stripe width.
From the detailed comparison with rather narrow chemical stripes (see \fref{fig:narrow}(b)) we infer that even for
such cases the Derjaguin approximation describes the actual behavior quite well, as expected from a corresponding
theoretical analysis within mean-field theory \cite{troendle:2010}.
We conclude that the quantitatively successful comparison between the experimental data and the theoretical predictions
demonstrates the significance and reliability of the critical Casimir effect for colloidal suspensions and reveals
a new means of using them as model systems in soft-matter physics or in applications in integrated
micro- or nano-devices.
\subsection{Perspectives}
\begin{quote}
  {\it ``After many failures, involving many fluid samples, I did succeed in levitating a hard platelet.''}\\
  \hspace*{20mm}--- Robert Evans.
\end{quote}
Since quantum-electrodynamic Casimir forces acting on the scale of nm and $\mu$m are typically always
attractive, they are responsible for stiction occurring in micro- and nano-mechanical
devices.
According to a generalization of the Earnshaw theorem, for quantum Casimir forces acting
on conventional materials in vacuum, stable levitation is not possible \cite{rahi:2010}.
Therefore, in order to overcome the problem of stiction using the quantum-electrodynamic Casimir effect
the system must be immersed in a fluid and the bodies have to be designed very specifically  \cite{munday:2009,rodriguez:2010}.
Obviously however, for a system immersed in a fluid  it is natural to use solvent-mediated forces such as the
critical Casimir forces to control its behavior.
Since one can obtain attractive and repulsive \emph{critical} Casimir forces by suitable surface
treatments as discussed here, a properly chosen combination of chemical substrate patterns 
may easily lead to levitation of a colloidal particle at a stable height above the substrate \cite{troendle:2010}.
One finds that this critical Casimir levitation not only can be achieved without involving any other additional force but
that the stable equilibrium distance between the wall and the colloid can be tuned steeply by varying the temperature
\cite{troendle:2010}.
Recently \cite{troendle:2010} it was predicted that the ranges of geometrical parameters for a pattern
of laterally alternating stripes (as discussed here), which allow for critical Casimir levitation of
colloidal particles, are rather large.
Moreover, critical Casimir levitation is robust even in the presence of other
forces and of thermal fluctuations \cite{troendle:2010} as they occur typically in colloidal suspensions, so
that an experimental realization of this effect appears to be possible.
These results show that chemical patterning of substrates allows one to design the critical Casimir effect over
a wide range of properties.
\par
\begin{quote}
  {\it ``Ions are the invention of the devil.''}\\
  \hspace*{20mm}--- Robert Evans.
\end{quote}
By adding salt to the solvent of the colloidal suspension the electrostatic
repulsion between the substrate and a colloid is strongly screened; this provides the
possibility to explore the spatial variation of critical Casimir forces over a much wider range of distances and
correlation lengths of the fluid.
Moreover, a rather complex behavior of the resulting forces acting on the colloid
can be expected because electrostatic forces induced by ions lead to rather delicate physical features.
Recently, preliminary measurements have shown a rich and novel behavior of a
colloid immersed in a salt-rich solvent upon approaching the critical point \cite{nellen:2010}.
The onset of strong attractive forces even several K below the critical temperature is observed and,
in addition, these forces are found to remain attractive, independent of the boundary
conditions, $(\pm,\pm)$ or $(+,-)$, throughout a large temperature range.
Currently it is not clear whether these observations can be entirely attributed to critical Casimir
forces or whether, e.g., the coupling between electrostatic interactions and critical
fluctuations has to be considered.
Thus, using these intriguing effects together with a chemical patterning of substrates
may open a new field of phenomena with potential applications in colloid science.
\section*{Acknowledgement}
It is a great pleasure for the authors to dedicate this paper to Bob Evans on the occasion of his 
65th birthday in view of his long lasting inspiration and encouragement across many fields of
statistical physics.
%

\label{lastpage}

\end{document}